\documentclass[aps,prc,groupedaddress,preprint
]{revtex4}
\usepackage{graphicx,bm}
\usepackage{amsmath,bm}
\DeclareMathOperator{\nrd}{\overleftrightarrow{\nabla}}

\begin{document}

\title{Short-range three-nucleon interaction from $A=3$ data
  and its hierarchical structure}

\author{L.\ Girlanda$^{\,{\rm a,b}}$, A.\ Kievsky$^{\,{\rm c}}$,
 M.\ Viviani$^{\,{\rm c}}$
 and L.E.~Marcucci$^{\,{\rm c,d}}$
}
\affiliation{
$^{\,{\rm a}}$\mbox{Dipartimento di Matematica e Fisica ``E.~De Giorgi'', Universit\`a del Salento, I-73100 Lecce, Italy}\\
$^{\,{\rm b}}$\mbox{INFN, Sezione di Lecce, I-73100 Lecce, Italy}\\
$^{\,{\rm c}}$\mbox{INFN, Sezione di Pisa, I-56127 Pisa, Italy}\\
$^{\,{\rm d}}$\mbox{Dipartimento di Fisica ``E.~Fermi'', Universit\`a di Pisa, I-56127 Pisa, Italy}\\
}
\date{\today}

\begin{abstract}
We construct accurate models of three-nucleon ($3N$) interaction  by fitting, in a hybrid phenomenological approach, the low-energy constants parametrizing the subleading $3N$ contact operators
to the triton binding energy, $n-d$ scattering lengths,  cross section and polarization observables of $p-d$ scattering at 2~MeV center-of-mass energy.
These models lead to a satisfactory description of polarized $p-d$ scattering data in the whole energy range below the deuteron breakup threshold. In particular,  the long-standing $A_y$ puzzle seems to be solved thanks to the new terms considered in the $3N$ force. Two types of hierarchies among the subleading contact operators are also derived, based on the large-$N_c$ counting and on a recently proposed relativistic counting. We test these hierarchies  against the same experimental data and show that they are respected at a reasonable level.
  
\end{abstract}

%\pacs{}
\index{}\maketitle

\section{Introduction}
Recent years have witnessed substantial progress in the development of accurate representations of the 
nuclear interaction, in both the two-nucleon ($NN$) and three-nucleon ($3N$) sectors \cite{Entem2017,Epelbaum2015,Reinert2017,Piarulli2015,Piarulli2016,Ekstroem2013,Carlsson2016,Krebs2013}. Particular emphasis 
has been put on the systematic framework provided by chiral effective field theory (ChEFT) \cite{Epelbaum2009,Machleidt2011}. 
The utility of chiral symmetry as organizing principle of the various components of the nuclear interaction 
depends on the convergence properties of the corresponding perturbation series, which reflect in turn the 
separation of the scales at which the nuclear interaction reveals its full complexity. In the $NN$ sector, chiral potentials, developed up to the 4th and 5th order of the low-energy expansion,  provide an extremely accurate description of the
$NN$ data up to laboratory energies of 300 MeV with a $\chi^2$  per degree of freedom ($\chi^2$/d.o.f.) close to one. The three-nucleon 
interaction (TNI) shows up in this framework as a small perturbation to the $NN$ interaction arising at the next-to-next-to-leading  order (N2LO), and depends only on two low-energy constants (LECs) up to following order N3LO \cite{Epelbaum2002,Bernard2008}.  After determining the two TNI LECs from two $3N$ data (usually they are the $^3$H binding energy and the doublet $n-d$ scattering length 
or tritium $\beta$-decay) the calculated $\chi^2$/d.o.f. of available low-energy $N-d$ observables takes values as large as several hundreds~\cite{Kievsky2001,Marcucci2009}. 
This well known fact regards unexplained discrepancies between theory and experiment in low-energy $N-d$ scattering, most notably 
in polarization observables of elastic scattering, as the so-called $A_y$ puzzle 
\cite{Kievsky1995,Golak2014,Viviani2013}. Attempts to trace back this problem to deficiencies in the
description of the low-energies $NN$ $p$-waves showed that it is impossible to simultaneously describe the low-energy $NN$ and $3N$ database using solely $NN$ forces~\cite{Entem2002}. Accordingly, these discrepancies
indicate a limited flexibility in the $3N$ force at the order considered.
To improve the description, further LECs, parametrising subleading contact terms contributing at N4LO, could 
be necessary. This would imply a slower convergence of the ChEFT series than expected, or the necessity of 
promoting short-range contact terms in the low-energy counting \cite{Nogga2005,Birse2006,Pavon2015,Kievsky2017}. In the present paper we 
focus on this component of the TNI to assess its relevance in the resolution of the above discrepancies.
The subleading TNI contact potential has been derived in Ref.~\cite{Girlanda2011}. It was shown that it consists
of 10 independent terms involving different combinations of the space-spin-isospin variables. 
Preliminary studies \cite{Girlanda2016} already indicated that the associated operatorial structures provide enough 
flexibility to improve the description of polarization observables in low-energy $N-d$ scattering.
In particular, assigning values to some of the accompanying LECs, it is possible to describe 
the two vector analyzing powers $A_y$ and $iT_{11}$ in good agreement with the experimental data. 
This preliminary study has opened the door to the possibility of fixing the TNI LECs from $3N$
scattering data.
%It should be noticed that a systematic approach of this kind was not accomplished yet.
In the present paper we intend to start a systematic study using $N-d$ scattering data 
to fix the 10 contact TNI LECs from a fitting procedure similar to what is done in the determination
of the $NN$ interaction. As a first step in this direction, and
following the previous analysis, we take the leading part of the force to be
the AV18 $NN$ potential~\cite{Wiringa1995}, with only the point-Coulomb interaction retained in the electromagnetic terms, in conjunction with the Urbana IX (UIX) model of 
TNI~\cite{Pudliner1997}. We fit the corresponding LECs to very precise $p-d$ cross section and polarization observables 
at  center of mass energy  $E_{\mathrm{cm}}=2$~MeV (or proton energy $E_p=3$~MeV)   \cite{Shimizu1995} for different choices of the contact short-distance cutoff $\Lambda$ between 
200 and 500~MeV. The resulting Hamiltonian is then used to predict the observables at other energies with
an overall satisfactory agreement inside the energy range explored.

On a more formal ground, we derive a hierarchy among these LECs as dictated by 't~Hooft large-$N_c$ limit 
of QCD \cite{thooft1974,Witten1979}. We also consider a recently proposed alternative counting for contact operators which 
does not rely on the non-relativistic expansion for nucleons \cite{Ren2018}, and classify the $3N$ 
contact operators appearing at the leading order in this counting. The simplified models for the contact TNI 
resulting from the leading orders of these schemes are also tested against the same experimental data, 
obtaining results of comparable quality. In particular, the relativistic counting seems to provide a natural 
explanation for a large spin-orbit term, as requested to explain the $A_y$ puzzle~\cite{Kievsky1999}. Strictly 
speaking these expansion schemes could only be tested in association with a chiral $NN$ potential derived in 
the same framework. However, we take the indications from the present ``hybrid'' approach as suggestive of 
their effectiveness.

The paper is organized as follows. In Section \ref{sec:model} we present our model of TNI interaction. Since 
the $p-d$ scattering can mostly probe the isospin $T=1/2$ component of the TNI we also discuss the projection of 
the model in this channel. There is also a $T=3/2$ component which we leave  undetermined: it could be fixed by other experimental 
observables. In Section \ref{sec:fit} we describe the variational procedure we use to solve the $p-d$ scattering 
problem, which is based on the expansion on the Hyperspherical Harmonics (HH method), and 
we describe the adopted fitting strategy and the results. The predictions at lower energies 
are compared to available experimental data in Section \ref{sec:pred}. In Section \ref{sec:hier} we determine the 
simplification of the subleading TNI implied by the large-$N_c$ limit and by the relativistic counting, 
and the corresponding test against experimental data. Finally, Section \ref{sec:concl} contains some concluding 
remarks. Details of the Fierz identities for covariant nucleon trilinears are collected in  Appendix~\ref{app:fierz}.

\section{The subleading TNI} \label{sec:model}
In Ref.~\cite{Girlanda2011}  all subleading $3N$ contact terms, compatible with the discrete symmetries of QCD and with the relativity constraints \cite{Girlanda2010}, were classified. Pauli principle severely reduces their number to only 10 independent structures.
From the Lagrangian density,
\begin{equation}
  {\cal L}_{3N} = - \sum_{i=1}^{10} E_i O_i,
\end{equation}
by appropriately choosing the momentum cutoff as dependent only on momentum transfers, an explicit representation  of the associated $3N$ potential can be derived, which is local in coordinate space and depends on a short-distance cutoff $\Lambda$ and the 10 subleading LECs $E_i$, $i=1,...,10$. It is explicitly written as 
\begin{eqnarray}
V^{(2)}=\sum_{i\neq j\neq k} && (E_1 + E_2 {\bm \tau}_i \cdot {\bm \tau}_j + E_3 {\bm \sigma}_i \cdot {\bm \sigma}_j + E_4 {\bm \tau}_i \cdot {\bm \tau}_j  {\bm \sigma}_i \cdot {\bm \sigma}_j)  \left[ Z_0^{\prime\prime}(r_{ij}) + 2 \frac{Z_0^\prime(r_{ij})}{r_{ij}}\right] Z_0(r_{ik})  \nonumber \\
&& + (E_5 +E_6 {\bm \tau}_i\cdot{\bm \tau}_j) S_{ij} \left[ Z_0^{\prime\prime}(r_{ij}) - \frac{Z_0^\prime(r_{ij})}{r_{ij}}\right] Z_0(r_{ik}) \nonumber \\
&& + (E_7 + E_8 {\bm \tau}_i\cdot{\bm \tau}_k) ( {\bf L}\cdot {\bm S})_{ij} \frac{Z_0^\prime(r_{ij})}{r_{ij}} Z_0(r_{ik}) \nonumber \\
&& + (E_9 + E_{10} {\bm \tau}_j \cdot {\bm \tau}_k) {\bm \sigma}_j \cdot \hat {\bf r}_{ij}  {\bm \sigma}_k \cdot \hat {\bf r}_{ik} Z_0^\prime(r_{ij}) Z_0^\prime(r_{ik})
\end{eqnarray}
where $S_{ij}$ and $ ( {\bf L}\cdot {\bm S})_{ij}$ are respectively the tensor and spin-orbit operators for particles $i$ and $j$, and the function $Z_0(r)$ is the Fourier transform of the cutoff function $F({\bf p}^2;\Lambda)$,
\begin{equation}
Z_0(r;\Lambda)=\int \frac{d {\bf p}}{(2 \pi)^3} {\mathrm{e}}^{i {\bf p}\cdot {\bf r}} F({\bf p}^2;\Lambda).
\end{equation}
We adopt the following choice for the cutoff function
\begin{equation}
  F({\bf p}^2,\Lambda)=\exp\left[-\left(\frac{{\bf p}^2}{\Lambda^2}\right)^2\right],
\end{equation}
which has the advantage of preserving the low-energy counting up to the order we are considering.

In this paper we consider a nuclear interaction consisting of the AV18 $NN$ potential the UIX TNI and an additional interaction given by
\begin{equation} \label{eq:vct}
V^{CT}_{3N}= V^{(0)} + V^{(2)},
\end{equation}
where the leading $3N$ contact potential $V^{(0)}$ is written as
\begin{equation}
V^{(0)} = \sum_{i\neq j\neq k} E_0 Z_0(r_{ij}) Z_0(r_{ik}).
\end{equation}

%\subsection{Isospin projection} \label{sec:isoproj}
Since  the deuteron is an isosinglet state, matrix elements between $N-d$ states only probe the total $T=1/2$ component of the TNI. In order to identify this component we use the projectors on the two isospin channels, which for the three-nucleon system take the form
\begin{equation}
P_{1/2}=\frac{1}{2} - \frac{1}{6} \left( {\bm \tau}_1\cdot {\bm \tau}_2 +{\bm \tau}_1\cdot {\bm \tau}_3 +{\bm \tau}_2\cdot {\bm \tau}_3 \right),
\end{equation}
and $P_{3/2}=1-P_{1/2}$.
The $3N$ potential $V^{(2)}$ can be expressed in momentum space as
\begin{equation}
  V^{(2)}=\sum_iE_i O_i,
\end{equation}
where the 10 $O_i$ operators are
\begin{equation}
\begin{array}{ll}
 O_1=-{\bf k}_i^2, & 
O_2=-{\bf k}_i^2 {\bm \tau}_i\cdot {\bm \tau}_j,\\
O_3=-{\bf k}_i^2 {\bm \sigma}_i\cdot {\bm \sigma}_j,&
O_4=-{\bf k}_i^2 {\bm \sigma}_i\cdot {\bm \sigma}_j {\bm \tau}_i\cdot {\bm \tau}_j \\
O_5=-3 {\bf k}_i\cdot {\bm \sigma}_i {\bf k}_i\cdot {\bm \sigma_j} + {\bf k}_i^2 {\bm \sigma}_i \cdot {\bm \sigma}_j, &
O_6=(-3 {\bf k}_i\cdot {\bm \sigma}_i {\bf k}_i\cdot {\bm \sigma_j} + {\bf k}_i^2 {\bm \sigma}_i \cdot {\bm \sigma}_j,)  {\bm \tau}_i\cdot {\bm \tau}_j,\\
 O_7=-\frac{i}{4} {\bf k}_i \times ({\bf Q}_i - {\bf Q}_j)\cdot ({\bm \sigma}_i + {\bm \sigma}_j), &
O_8=-\frac{i}{4} {\bf k}_i \times ({\bf Q}_i - {\bf Q}_j)\cdot ({\bm \sigma}_i + {\bm \sigma}_j) {\bm \tau}_j\cdot {\bm \tau}_k, \\
O_9=-{\bf k}_i\cdot {\bm \sigma}_i {\bf k}_j\cdot {\bm \sigma}_j, & 
O_{10}=-{\bf k}_i\cdot {\bm \sigma}_i {\bf k}_j\cdot {\bm \sigma}_j {\bm \tau}_i\cdot {\bm \tau}_j,
\end{array}
\end{equation}
with ${\bf k}_i={\bf p}_i-{\bf p}_i'$, ${\bf Q}_i ={\bf p}_i + {\bf p}_i'$ and ${\bf p}_i$ (${\bf p}_i'$) the initial (final) momentum of the $i$-th nucleon, and a sum over $i\neq j\neq k$ is understood. The projections over isospin $T=1/2$, $(O_i)_{1/2} = P_{1/2} O_i P_{1/2}$ are given, using the relations derived in Ref.~\cite{Girlanda2011}, by
\begin{eqnarray}
(O_1)_{1/2}&=&O_1-\frac{1}{3} O_2 + \frac{1}{3} O_3 + \frac{1}{9} O_4 + \frac{1}{3} O_5  + \frac{1}{9} O_6 + 4 O_7 +\frac{4}{3} O_8 + O_9 + \frac{1}{3} O_{10},\\
(O_2)_{1/2}&=&\frac{2}{3} O_2 + \frac{1}{3} O_3 + \frac{1}{9} O_4 + \frac{1}{3} O_5  + \frac{1}{9} O_6 + 4 O_7 +\frac{4}{3} O_8 + O_9 + \frac{1}{3} O_{10},\\
(O_i)_{1/2}&=& O_i,\quad i=3,...,8\\
(O_9)_{1/2}&=& \frac{1}{6} O_2 -\frac{1}{6} O_3 -\frac{1}{18} O_4 - \frac{1}{6} O_5 -\frac{1}{18} O_6 - 2 O_7 - \frac{2}{3} O_8 + \frac{1}{2} O_9 -\frac{1}{6} O_{10},\\
(O_{10})_{1/2}&=& \frac{1}{6} O_2 -\frac{1}{6} O_3 -\frac{1}{18} O_4 - \frac{1}{6} O_5 -\frac{1}{18} O_6 - 2 O_7 - \frac{2}{3} O_8 - \frac{1}{2} O_9 +\frac{5}{6} O_{10}.
\end{eqnarray}
By examining the above relations, we find that there are 9 purely $T=1/2$ combinations, e.g.
\begin{equation} \label{eq:t12}
  (O_1 - O_2), \quad (O_2 + 2 O_{10}),  \quad O_{i=3,...,8},  \quad (O_9 - O_{10}),
\end{equation}
and  a single purely $T=3/2$ combination of operators, e.g.
\begin{equation} \label{eq:t32}
O_{3/2}=3 O_2 - 3 O_3 - O_4 - 3 O_5 - O_6 - 36 O_7 - 12 O_8 - 9 O_9 - 3 O_{10}.
\end{equation}
Notice that, in order to derive the above projections, Fierz transformations have been repeatedly used. Therefore the conclusion only holds up to cutoff effects: indeed the cutoff smears the contact interactions and, as a consequence, the three nucleons, no longer at the same position, are much less constrained by the Pauli principle.
Thus, only 9 combinations of LECs may enter $p-d$ observables, and no full determination of all the 10 LECs will be possible without adding an extra  $T=3/2$ observable.
We may as well start from a Hamiltonian written in terms of the isospin-projected operators Eqs.~(\ref{eq:t12})-(\ref{eq:t32}) with LECs $h_i$, $i=1,...,9$ and $h_{3/2}$ respectively, in one-to-one correspondence with the $E_i$, e.g.
\begin{equation}
  h_{3/2}=\frac{1}{18} \left[ 2 \left( E_1+E_2 \right) - E_9 - E_{10} \right].
  \end{equation}
Dropping the $T=3/2$ operator from the Hamiltonian, which does not affect the $p-d$ observables, amounts to setting $h_{3/2}=0$, leading to the relation
\begin{equation} \label{eq:t12constraint}
  2 (E_1 + E_2) - E_9 - E_{10} =0.
\end{equation}
Thus we may effectively impose the above constraint when fitting to $p-d$ observables, and shifting all the LECs by an amount proportional to the $T=3/2$ LEC multiplying $O_{3/2}$ according to Eq.~(\ref{eq:t32}), once we add this extra observable.

\section{Numerical determination of the contact LECs} \label{sec:fit}
We use the HH method to solve the 3-body Schroedinger equation, as reviewed in Ref.~\cite{Kievsky2008}. The $N-d$ scattering wave function, below the deuteron breakup threshold, is written as the sum of an internal and an asymptotic part,
\begin{equation}
\Psi_{LSJJ_z} = \Psi_{\mathrm{C}} + \Psi_{\mathrm{A}},
\end{equation}
where the internal part is expanded in Hyperspherical Harmonics,
\begin{equation} \label{eq:psicore}
\Psi_C = \sum_\mu c_\mu \Phi_\mu ,
\end{equation}
$\mu$ denoting a set of quantum numbers necessary to completely specify the basis element, while the asymptotic 
part, $\Psi_{\mathrm{A}}$, describes the relative motion between the nucleon and the deuteron at large separation, which takes the form 
of a linear combination of the regular and irregular solutions of the free (or Coulomb) $N-d$ Schroedinger equation 
at relative momentum $q$ (corresponding to energy $E$), duly regulated at small distance. Therefore, denoting these solutions with
$\Omega^\lambda_{LSJJ_z}$,  $\lambda=R,I$ respectively, we can write,
\begin{equation}
\Psi_A^{LSJJ_z}=\Omega^R_{LSJJ_z} + \sum_{L'S'}{\cal R}^J_{LS,L'S'}(q) \Omega^I_{L'S'JJ_z}.
\end{equation}
The weights ${\cal R}^J_{LS,L'S'}$ of the irregular solution relative to the regular one are the $K$-matrix
elements. It is related to the $S$-matrix from the relation $S=(1+iK)(1-iK)^{-1}$. The $K$-matrix,
that determines the scattering phase shifts and mixing parameters, together with the coefficient $c_\mu$ in 
Eq.~(\ref{eq:psicore}) are obtained from the Kohn variational principle. The principle can be formulated
in its real or complex form~\cite{Kievsky1997} and requires that the functional 
\begin{equation} \label{eq:kohn}
\left[ {\cal R}^J_{LS,L'S'}(q) \right] = {\cal R}^J_{LS,L'S'}(q) - 
\langle \Psi_{L'S'JJ_z} | H - E | \Psi_{LSJJ_z} \rangle
\end{equation}
be stationary under changes of the variational parameters in $\Psi_{LSJJ_z}$, with the asymptotic part normalized 
such that
\begin{equation}
\langle \Omega^R_{LSJJ_z}|H-E|\Omega^I_{LSJJ_z} \rangle -\langle \Omega^I_{LSJJ_z} |H-E |\Omega^R_{LSJJ_z}\rangle =1.
\end{equation}  
This implies that the weights ${\cal R}_{LS,L'S'}^J$ must solve the linear system
\begin{equation}
\sum_{\tilde L \tilde S} {\cal R}^J_{LS,\tilde L \tilde S} X_{L'S',\tilde L \tilde S} = Y_{LS,L'S'}
\end{equation}
where
\begin{equation}
X_{LS,L'S'}=\langle \Omega^I_{LSJJ_z} + \Psi_C^I| H-E|\Omega^I_{L'S'JJ_z}\rangle, \quad
Y_{LS,L'S'}=-\langle \Omega^I_{LSJJ_z} + \Psi_C^R| H-E|\Omega^I_{L'S'JJ_z}\rangle, 
\label{eq:asympt}
\end{equation}
and the internal functions $\Psi_C^{\lambda}$ have coefficients $c_\mu^{\lambda}$ solutions of
\begin{equation} \label{eq:linmatr}
\sum_{\mu'} \langle \Phi_\mu|H-E|\Phi_{\mu'}\rangle c_{\mu'}^\lambda= 
- \langle \Phi_\mu|H-E| \Omega_{LSJJ_z}^{\lambda} \rangle,
\end{equation}
with $\lambda=R,I$.
A second-order estimate is then obtained by substituting the obtained weights ${\cal R}^J_{LS,L'S'}$ 
into Eq.~(\ref{eq:kohn}). From Eqs.~(\ref{eq:asympt}) and (\ref{eq:linmatr}) we notice that, in order 
to solve the linear problem, the matrix elements of the Hamiltonian $H$ have to be computed between the 
HH basis elements and the asymptotic functions. Decomposing the Hamiltonian as
\begin{equation} \label{eq:lham}
H=T+V= T+V_{2N}+V_{3N}=H_L+V^{(0)}+V^{(2)}\; ,
\end{equation}
where $H_L$ is the leading Hamiltonian containing the kinetic energy $T$ plus the selected
two- and three-body force and $V^{(0)}+V^{(2)}$ are the leading and subleading contact interactions,
the linear system of Eq.~(\ref{eq:linmatr}) results
\begin{equation} \label{eq:linmatr2}
\sum_{\mu'} c_\mu^{\lambda} \langle \Phi_\mu|H_L+\sum_{i=0,10}E_iV_i-E|\Phi_{\mu'}\rangle = 
- \langle \Phi_\mu|H_L+\sum_{i=0,10}E_iV_i-E| \Omega_{LSJJ_z}^{\lambda} \rangle,
\end{equation}
which  can be put in the matricial form
\begin{equation} \label{eq:linmatr3}
\sum_{\mu'} \big[ (H_L)_{\mu\mu'}+\sum_{i=0,10}E_i (V_i)_{\mu\mu'}-E N_{\mu\mu'}\big] c_{\mu'}^\lambda = 
- (H_L)_{\mu\lambda}+\sum_{i=0,10}E_i(V_i)_{\mu\lambda}-EN_{\mu\lambda}, 
\end{equation}
where $(H_L)_{\mu\mu'}$ denote the matrix elements of $H_L$ between the corresponding basis states and similarly for the other operators. Here $E_i,V_i$ are the contact leading ($i=0$) and subleading ($i=1,\ldots,10$) LECs and
operators respectively. As can be seen the problem has been reduced to a linear one: 
the contact potential energy can be computed as a linear combination of several matrices, 
one for $V^{(0)}$ and one for each operator appearing in $V^{(2)}$. These matrices 
can be computed once for all, weighted by the corresponding LECs. 
Using the Kohn variational principle in the complex formalism, a particular set of LECs can be used to compute the corresponding $S$- or $T$-matrix for each $J^\pi$
state from which the observables at a particular energy $E$ can be obtained.
To this end we calculate the $N-d$ transition matrix
$M$ decomposed as a sum of the Coulomb amplitude $f_c$ plus
a nuclear term
\begin{eqnarray}
M^{SS'}_{\nu\nu'}(\theta)& = &f_c(\theta)\delta_{SS'}\delta_{\nu\nu'}+
{\sqrt{4\pi}\over k}\sum_{L,L',J}\sqrt{2 L+1}(L0S\nu|J\nu)(L'M'S'\nu'|J\nu) 
\nonumber \\
&& \,\exp[i(\sigma_L+\sigma_{L'}-2\sigma_0)]\;
 T^J_{LS,L'S'} \; Y_{L'M'}(\theta,0) \;\; ,
\label{tm}
\end{eqnarray}
where the matrix $M$ is a $6\times6$ matrix corresponding to the couplings of the
spin $1$ and spin $1/2$, of the deuteron and third particle, 
to $S,S'=1/2$ or $3/2$ with projections $\nu$ and $\nu'$.
The quantum numbers $L,L'$ are the relative orbital angular momentum between
the deuteron and the third particle and $J$ is the total angular momentum
of the three-nucleon state. The matrix elements $T^J_{LS,L'S'}$ form
the $T$-matrix of a Hamiltonian containing the nuclear plus Coulomb interactions,
$\sigma_L$ are the Coulomb phase--shifts. The $n-d$ case is recovered with $f_c=\sigma_L=0$. 

Let us first determine the expected sizes of the LECs which,
according to na\"ive dimensional analysis \cite{Manohar1984,Georgi1993}, are as follows,
\begin{equation}
E_0 \sim \frac{1}{F_\pi^4 \Lambda}, \quad E_i \sim \frac{1}{F_\pi^4 \Lambda^3}, \quad i=1,...,10,
\end{equation}
where $F_\pi=92.4$~MeV is the pion decay constant and $\Lambda$ is the hadronic scale. This counting is 
expected in the pionful theory. In the pionless case the LECs may also receive contributions from virtual 
pion exchanges, which will produce extra factors of $\Lambda^2/M^2_\pi$. We therefore extract physical 
dimensions and write
\begin{equation}
E_0 = \frac{e_0}{F_\pi^4 \Lambda},\quad E_i = \frac{e_i}{F_\pi^4 \Lambda^3}, \quad i=1,...,10,
\end{equation}
with $e_0 \sim e_i \sim O(1)$ if natural.

In the determination of the LECs we make use of the
following data: the triton binding energy, the doublet and quartet $n-d$ scattering lengths \cite{Schoen2003,Dilg1971}
and several $p-d$ scattering observables at 3~MeV proton energy for which a very precise set of data
exists~\cite{Shimizu1995}. The $p-d$ observables used for the fit are the differential cross section, 
the two vector analyzing powers $A_y$ and $iT_{11}$ and the three tensor analyzing powers 
$T_{20}$, $T_{21}$ and $T_{22}$. For each choice of the 10 subleading LECs, subjected to the $T=1/2$ constraint (\ref{eq:t12constraint}), 
we redetermine the leading contact LEC $E_0$ from the experimental triton binding energy. We then fit the 
experimental doublet and quartet $N-d$ scattering length \cite{Schoen2003,Dilg1971} and the six $p-d$ scattering observables 
at $E_p=3$~MeV~\cite{Shimizu1995}, amounting to $\sim300$ experimental data.
The theoretical observables are calculated solving
Eq.~(\ref{eq:linmatr3}) for a set of $E_i$ coefficients for different $J^\pi$ states. The obtained $S$-matrix  
(or $T$-matrix) is used to calculate the transistion matrix $M$ from which the observables are directly
calculated~\cite{Gloeckle1996}. At the energy considered, states up to $L=2$ are calculated using
the full Hamiltonian whereas for $L>2$ only the two-body potential was included up to a maximum value
of $L=6$.

For the differential cross section we include in 
the $\chi^2$ an overall normalization factor $Z$ of the data points,
\begin{equation}
  \chi^2 =\sum_i \frac{(d_i^{\mathrm{exp}}/Z-d_i^{\mathrm{th}})^2}{(\sigma_i^{\mathrm{exp}}/Z)^2},
\end{equation}
with $Z$ obtained from the minimization condition as
\begin{equation}
  Z=\frac{\sum_i d_i^{\mathrm{exp}} d_i^{\mathrm{th}}/(\sigma_i^{\mathrm{exp}})^2}{\sum_i  (d_i^{\mathrm{th}})^2/(\sigma_i^{\mathrm{exp}})^2},
\end{equation}
and checked that $Z$ never differs from 1 by more than 2\%~\cite{Kievsky2001b}.
For the other observables, we treat the normalization $Z=1.00\pm 0.01$ as an experimental datum, to be added to the $\chi^2$, since, according to Ref.~\cite{Shimizu1995} the systematic uncertainty is estimated as 1\%.

\begin{table}
  \begin{tabular}{|c|c|c|c|c|}
    \hline
$\Lambda$ (MeV)    & 200&300 &400 &500
    \\
    \hline
    $\chi^2$/d.o.f. & 1.7 & 1.7 & 1.7 & 1.7 \\
    $e_0$ & -1.174 & -6.193 &-4.674 & -2.910 \\
    $e_1$ & 2.734 & 1.889&-3.599 & -2.691 \\
    $e_2$ & -0.532& -0.428 &2.711 &1.716 \\
    $e_3$ & -2.334 &-3.311 &1.472 &1.762  \\
    $e_4$ & 2.227 &3.745 &-1.044 & -1.367 \\
    $e_5$ & -0.197 &0.434 &-0.485 &-1.134 \\
    $e_6$ & 0.682 &-0.413 &-0.854 &-1.079 \\ 
    $e_7$ & 7.450 &4.353 &1.326 &0.815 \\
    $e_8$ &2.235 &0.811 &-0.170 &-1.097 \\
    $e_9$ & 4.034 &0.969 &-1.195 &-1.316 \\
    $e_{10}$ & 0.370 &1.953 &  -0.582 &-0.633 \\
%    $B(^3$H) (MeV) & 8.482 & 8.482 & 8.482 &8.482 
\hline
    $a_2$ (fm) & 0.635 & 0.641 & 0.645 &0.650 \\ 
    $a_4$ (fm) & 6.32 & 6.32 &6.32 &6.32\\
    \hline
  \end{tabular}
\caption{$\chi^2$/d.o.f., values of the rescaled LECs $e_i$, $i=0,10$ and
predicted values of the doublet and quartet scattering lengths $a_2$ and $a_4$ for the four
cutoff considered. Note that  the doublet and quartet $n-d$ scattering lengths $a_2$ and $a_4$ are fitted to the experimental values $a_2=(0.645\pm0.003\pm0.007)$~fm \cite{Schoen2003} and $a_4= (6.35 \pm 0.02)$~fm \cite{Dilg1971}.}
\label{tab:uix}
\end{table}

\begin{figure}
  \centerline{\includegraphics[width=15cm]{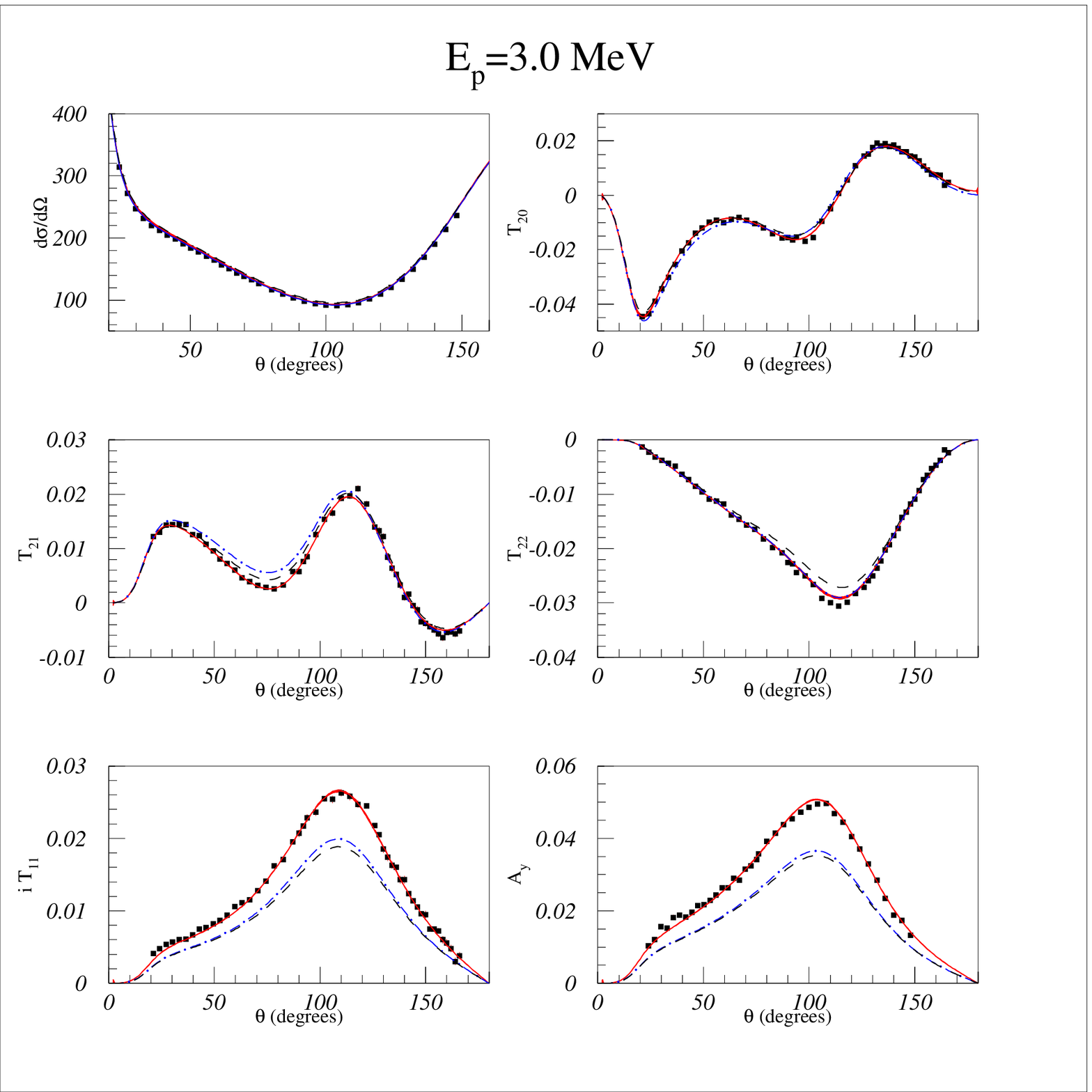}}
  \caption{Fit results to a set of cross section and polarization $p-d$ observables at 3~MeV proton energy \cite{Shimizu1995}, 
for $\Lambda=200-500$~MeV (red bands) as compared to the purely two-body AV18 interaction (dashed black lines) and to the AV18+UIX 
two- and three-nucleon interaction (dashed-dotted, blue lines). \label{fig:fit}}
\end{figure}

For a given initial set of LEC values we use Eq.~(\ref{eq:linmatr3}) to solve the scattering problem
and calculate the corresponding observables. Using the POUNDerS algorithm~\cite{Munson} we start an iterative
procedute to minimize the global $\chi^2$/d.o.f. of the data set description. After several iterations
the numerical procedure converge to a local minimum. We repeat the procedure using different initial
input of values trying to localize the deepest mimimum.
The results for the renormalized leading LEC $e_0$ and subleading LECs $e_i$, $i=1,10$ are displayed 
in Table~\ref{tab:uix}. The first line of the table shows the $\chi^2/$d.o.f. produced after the fit
and in the last two rows the predicted values for the doublet and quartet scattering lengths $a_2$ and $a_4$ 
are given. In all cases the triton binding energy of $8.482$ MeV is correctly described. 
Values of $\chi^2/$d.o.f. $\lesssim $1.7, 
are obtained for all  values of the short-distance cutoff between $\Lambda=200$ and 500~MeV. We show in Fig.~\ref{fig:fit} the fitted curves which, for each observable, form a (red) narrow band representing
the variation with the cutoff $\Lambda$ between 200 and 500~MeV. In addition, and for the sake of
comparison, we show the predictions using the two-body interaction AV18 and the AV18+UIX model. The underprediction
of $A_y$ and $iT_{11}$ are well visible in these two cases.
We also notice that the resulting TNI is not a small perturbation, as compared to the UIX. For instance, the contact terms $E_0,...,E_{10}$ contribute an overall attraction of the order of 1 MeV in the triton binding energy, which is also the result of partial cancellations among the different terms. Nevertheless, it is clear that  the subleading contact interaction terms have enough flexibility to improve
the description of the observables at this energy producing a very acceptable $\chi^2$/d.o.f. below 2.

\section{Predictions at lower energies} \label{sec:pred}
\begin{figure}
  \centerline{\includegraphics[width=15cm]{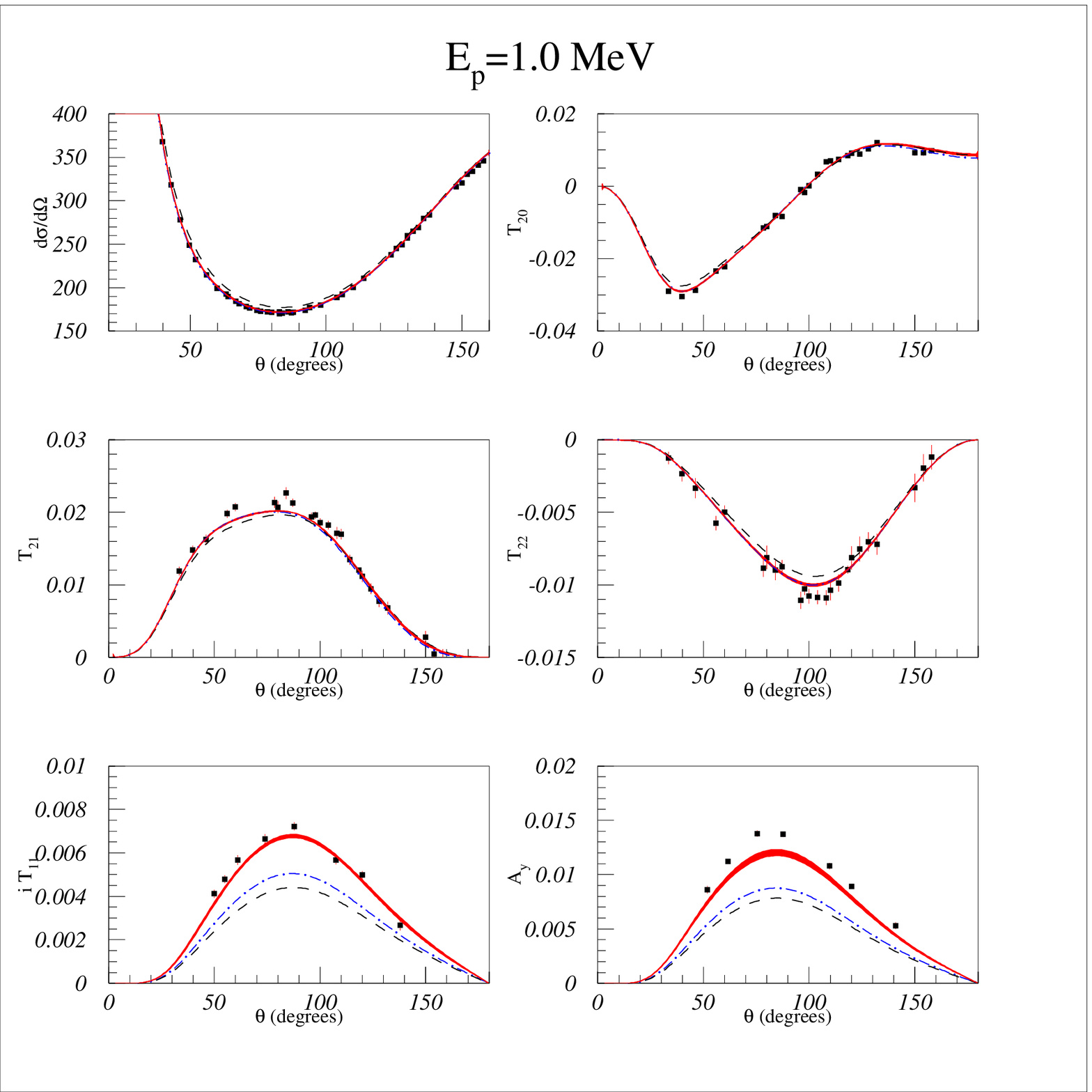}}
  \caption{Predictions of the models determined at $\Lambda$ between 200 and 500~MeV (red bands) for a set of cross section and polarization $p-d$ 
observables at $E_p=1$~MeV  energy, as compared to the  purely two-body AV18 interaction (dashed, black lines) and to the 
AV18+UIX two- and three-nucleon interaction (dashed-dotted, blue lines), in comparison to experimental data \cite{Wood2002}.  
\label{fig:e0_0.6}}\end{figure}

\begin{figure}
  \centerline{\includegraphics[width=15cm]{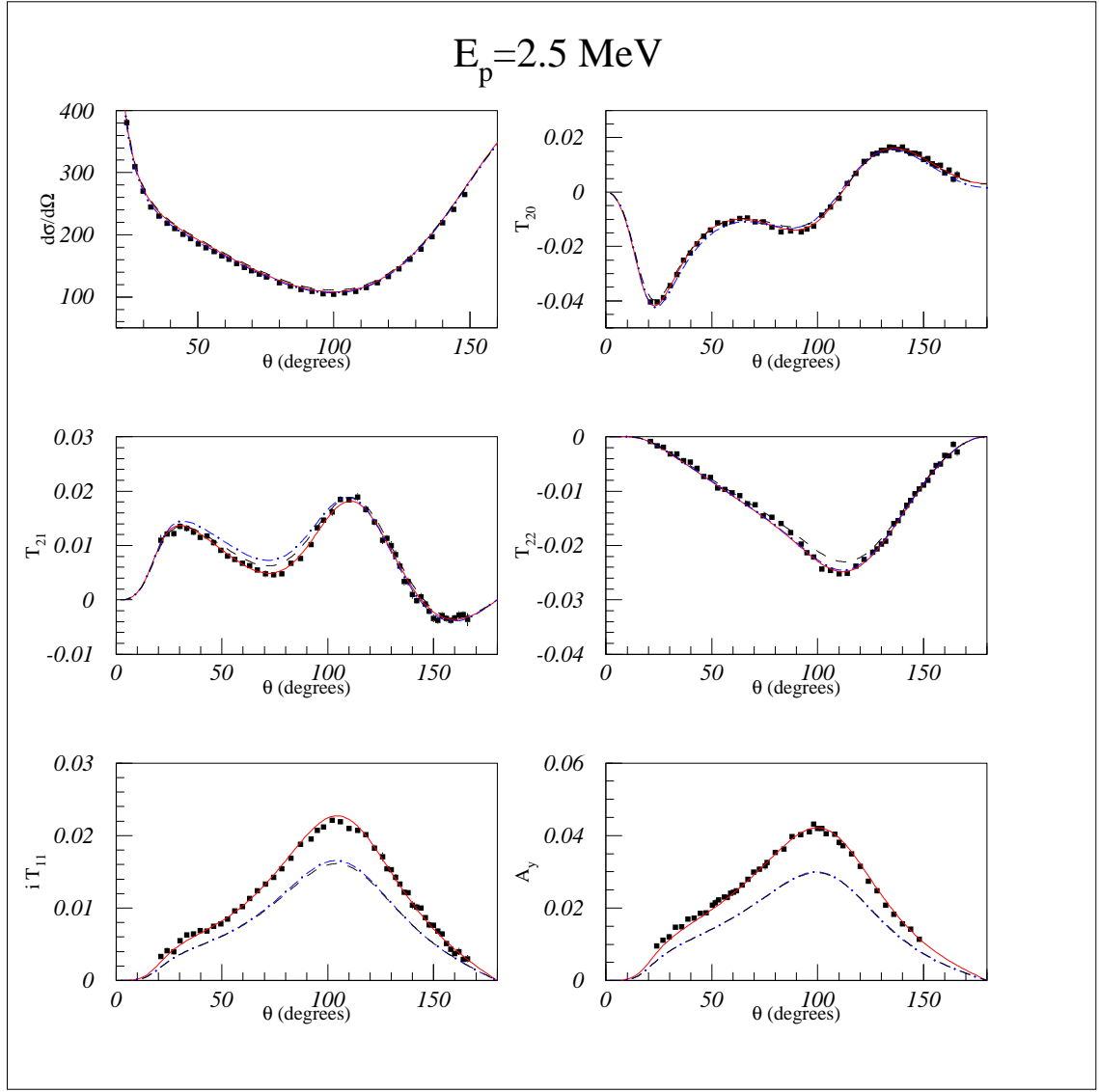}}
  \caption{Predictions of the model determined at $\Lambda=300$~MeV for a set of cross section and polarization $p-d$ 
observables at $E_p=2.5$~MeV proton energy (solid, red lines) as compared to the  purely two-body AV18 interaction (dashed, black lines) and to the
AV18+UIX two- and three-nucleon interaction (dashed-dotted, blue lines), in comparison to experimental data \cite{Shimizu1995}.  
\label{fig:e0_1.6}}
\end{figure}

\begin{figure}
  \centerline{\includegraphics[width=15cm]{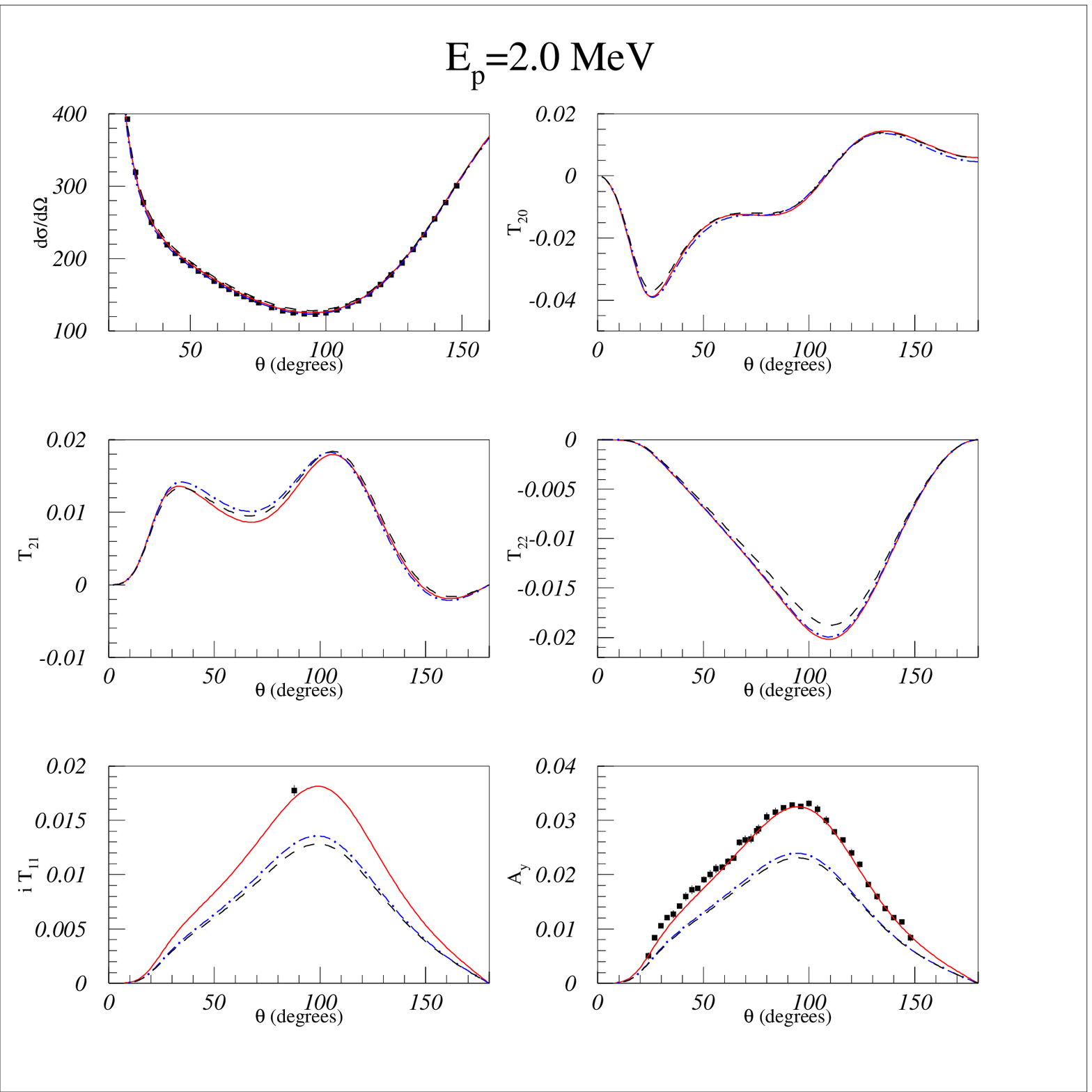}}
  \caption{Same as Fig.~\ref{fig:e0_1.6} but for $E_p=2$~MeV proton energy.  Data are from Ref.~\cite{Wood2002}.  
\label{fig:e0_1.3}}\end{figure}

\begin{figure}
  \centerline{\includegraphics[width=15cm]{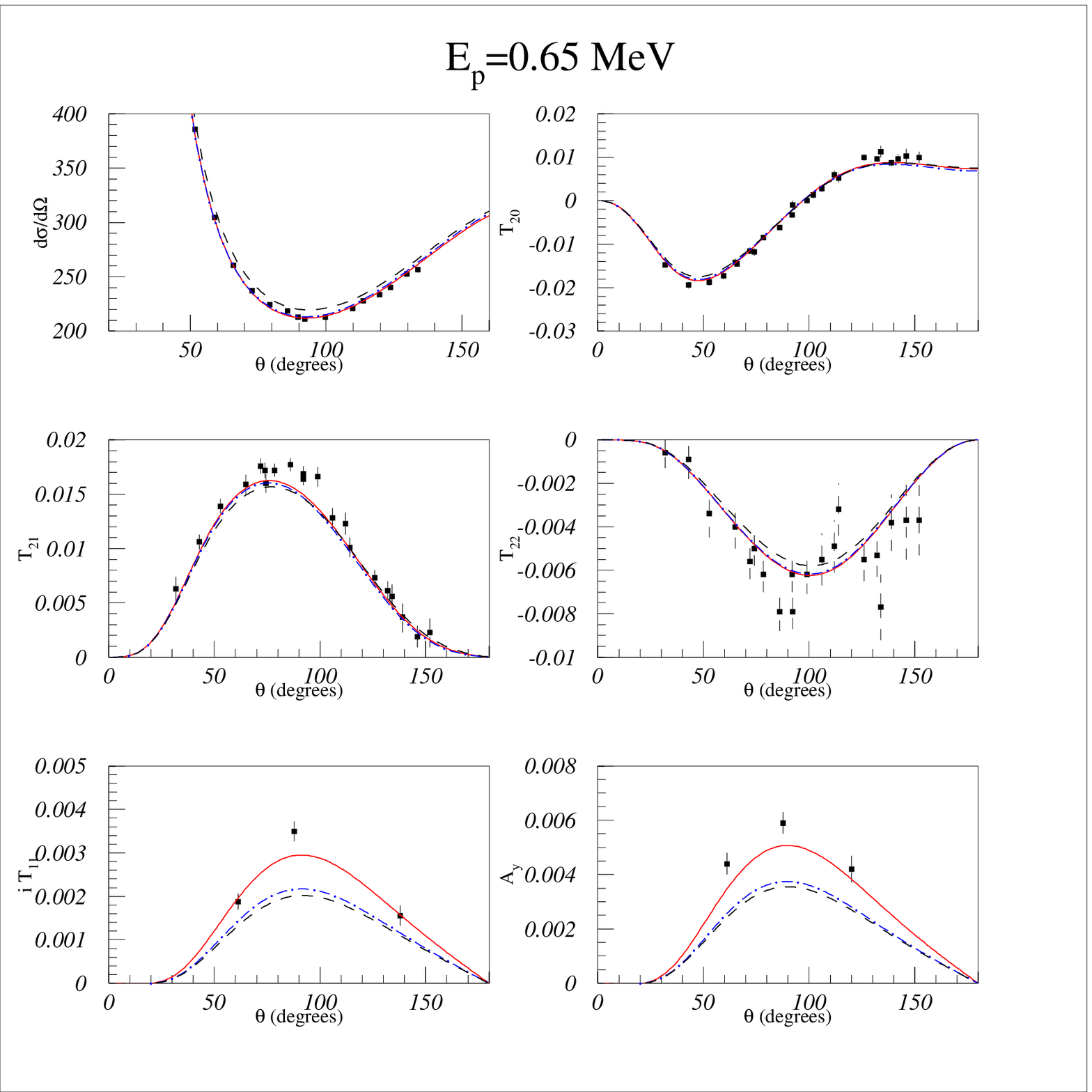}}
  \caption{Same as Fig.~\ref{fig:e0_1.6} but for proton energy $E_p=0.647$~MeV. Data are from Ref.~\cite{Brune2001}.  
\label{fig:e0_0.4}}\end{figure}

With the LECs determined at a proton energy of $E_p=3$~MeV, we can predict observables at other energies. 
To this end we select $p-d$ scattering below the deuteron breakup, and postpone the 
analysis at energies above the breakup as well as an energy-dependent fit to a forthcoming study. 
To make this analysis we choose the model corresponding to $\Lambda=300$~MeV. We have checked the cutoff dependence at proton energy $E_p=1.0$~MeV and found that it is small between 200 and 500~MeV, as shown in Fig.~\ref{fig:e0_0.6}.
Several observables have been measured at proton energies of $E_p=1.0$, 2.5, 2.0, 0.647~MeV 
\cite{Shimizu1995,Wood2002,Brune2001}. The theoretical predictions using the subleading contact interaction determined
with the fit at $E_p=3$~MeV are shown in Figs.~\ref{fig:e0_0.6}, \ref{fig:e0_1.6}, \ref{fig:e0_1.3}  and \ref{fig:e0_0.4} 
respectively.

\begin{figure}
  \centerline{\includegraphics[width=15cm]{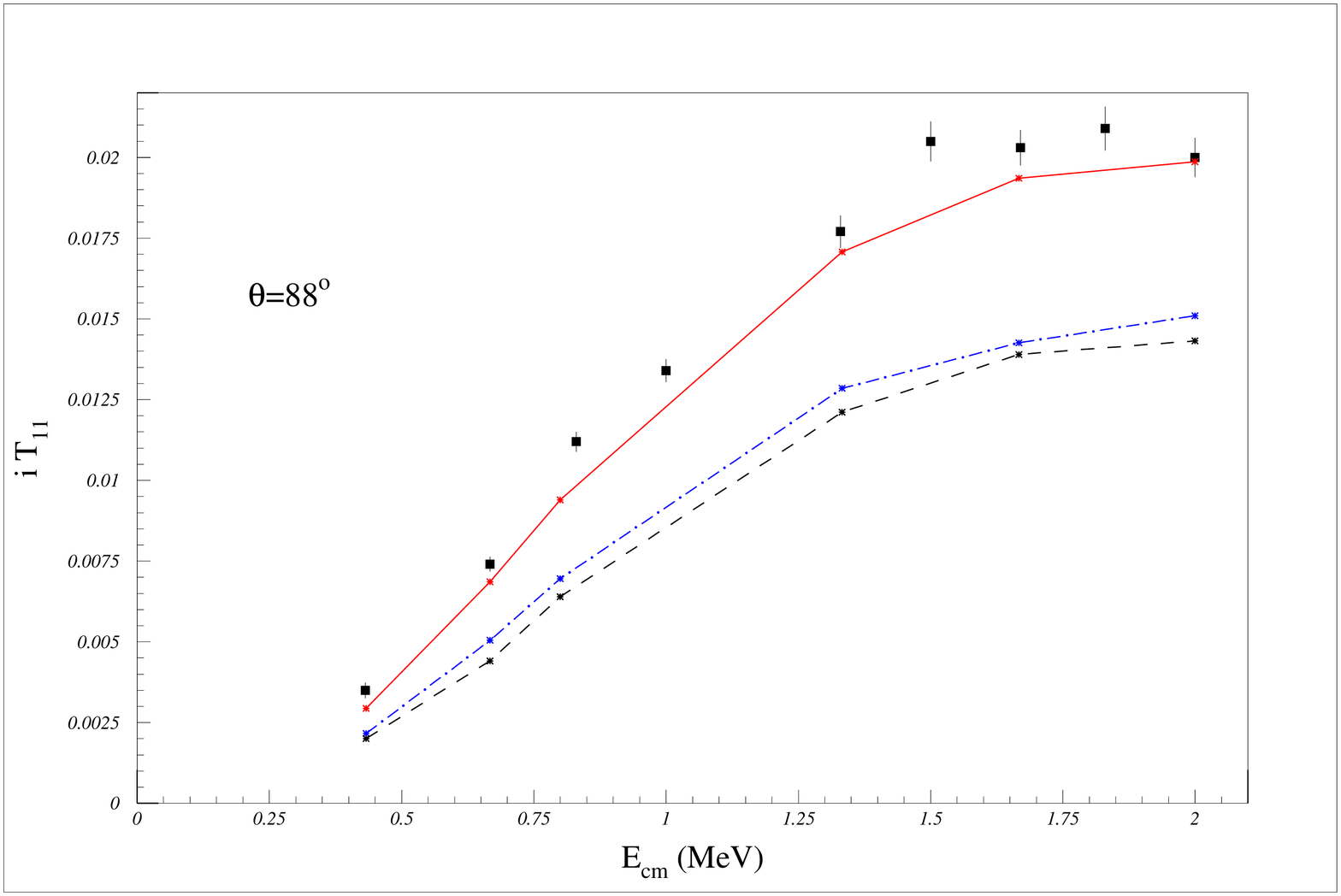}}
  \caption{Prediction of the model determined at $\Lambda=300$~MeV (red stars connected by solid lines) for the center of mass energy dependence of the vector 
polarization observable $i T_{11}$ at $\theta=88^\circ$ as compared to available experimental data of 
Ref.~\cite{Brune2001}. Also shown is the prediction corresponding to the purely two-body AV18 interaction (black stars connected by dashed lines) and to the AV18+UIX nuclear interaction (blues stars connected by dashed-dotted lines). 
\label{fig:exct11}}
  \end{figure}

By inspection of the  figures we observe an overall good agreement between theory and experiment for the observables, in particular the energy dependence of the 
analyzing power $A_y$ and the $i T_{11}$ observable is correctly encoded in the adopted contact interaction, although a small underprediction is still observed at the lowest energies. We have also investigated in Fig.~\ref{fig:exct11} the $i T_{11}$ scattering observable at  angle $\theta=88^\circ$, for which experimental data ara available in the intermediate energy domain.
From the figure we see that below $E_p=3$~MeV, corresponding to center of mass energy $E_{\mathrm{cm}}=2$~MeV, the observable is systematically slightly underpredicted by the model. 
A simultaneous fit at different energies will presumably lead to a more accurate description in the very low-energy domain. The investigation of this possibility is left for future work.

\section{Study of hierarchical structures in the subleading TNI} \label{sec:hier}
\subsection{Large-$N_c$ constraints}
 \label{sec:largenc}
 It is interesting to explore whether the proposed model of TNI fulfills the hierarchy dictated by the large-$N_c$ limit of QCD, thus revealing a closer connection to the underlying theory of the strong interactions. Indeed, in the 't Hooft limit \cite{thooft1974,Witten1979} where the number of colors $N_c\to \infty$ and the strong coupling constant $g$ scales like $g^2 N_c \sim 1$, from the scaling of connected baryon-baryon amplitudes it is possible to derive the large-$N_c$ scaling of nuclear potentials \cite{Kaplan1997,Kaplan1996}. Such scheme has proven to be qualitatively successful in the $NN$ case, providing in fact an expansion in $1/N_c^2$, and has been applied to the $3N$ case \cite{Phillips2013} as a guide to reduce the large number of operatorial structures.
In particular, one finds that the scaling of the two-nucleon ${\bm \sigma}\otimes {\bm \sigma}$ and ${\bm \tau} \otimes {\bm \tau}$ operators is twice suppressed with respect to ${\bf 1}\otimes {\bf 1}$ and ${\bm \sigma \tau}\otimes {\bm \sigma \tau}$, i.e.
\begin{equation} \label{eq:largenc0}
{\bm \sigma} \otimes {\bm \sigma} \sim {\bm \tau} \otimes {\bm \tau} \sim \frac{1}{N_c},\quad {\bf 1} \otimes {\bf 1} \sim {\bm \sigma \tau} \otimes {\bm \sigma \tau} \sim N_c.
\end{equation}
This is in phenomenological agreement with the size of the two LECs, $C_S$ and $C_T$, corresponding to leading order $NN$ contact interactions,
\begin{equation} \label{eq:nncontact}
{\cal L}_{NN}= -\frac{1}{2} C_S (N^\dagger N)^2 - \frac{1}{2} C_T(N^\dagger {\bm \sigma} N)^2,
\end{equation}
where $|C_S| \gg |C_T|$. Notice that the most general leading-order contact Lagrangian involving spin-isospin-1/2 baryons contains in principle four different operators,
\begin{equation}
{\cal L}_{NN}= c_1 (N^\dagger N)^2 + c_2 (N^\dagger {\bm \sigma} N)^2 + c_3 (N^\dagger {\bm \tau} N)^2 + c_4 (N^\dagger {\bm \sigma \tau} N)^2 \equiv \sum_i c_i o_i,
\end{equation}
which are related through Fierz-like identities by
\begin{equation} \label{eq:pauli}
o_3=-o_2-2 o_1,\quad o_4=-3 o_1.
\end{equation}
These relations do not conform with the large-$N_c$ scaling (\ref{eq:largenc0}). We observe however that the counting arguments which lead to the large-$N_c$ scaling never use the fact that the baryons are identical fermions. In particular, the same scaling would apply to scattering of distinguishable baryons. In this case we would have $c_1\sim c_4 \sim O(N_c)$ while $c_2 \sim c_3 \sim O(1/N_c)$. The indistinguishability of nucleons implies relations (\ref{eq:pauli}), which in turn allow to cast the effective lagrangian in the form (\ref{eq:nncontact}) with
\begin{equation}
C_S= -2 (c_1 - 2 c_3 - 3 c_4), \quad C_T = -2 (c_2 - c_3),
\end{equation}
whence the conclusion on the relative size of $C_S$ and $C_T$.
We thus learn that one way to implement the Pauli principle in the large-$N_c$ counting is to start with a redundant set of operators, establish the counting of the corresponding LECs, and impose the Pauli principle constraints afterwards.

In the notation of Table I of Ref.~\cite{Girlanda2011} there are 13  leading operators in the large-$N_c$ limit \cite{Phillips2013}, so that the $3N$ contact Lagrangian only depends on the 13 corresponding leading LECs,
\begin{eqnarray}
  {\cal L}_{3N}&=&c_{33} o_{33} + c_{37} o_{37} + c_{40} o_{40} + c_{43} o_{43} + c_{47} o_{47} \nonumber \\
  &&+c_{51} o_{51} + c_{55} o_{55} + c_{59} o_{59} + c_{60} o_{60} + c_{61} o_{61} + c_{62} o_{62} + c_{63} o_{63} +c_{64} o_{64}.
\end{eqnarray}
Using the Fierz identities obtained in Ref.~\cite{Girlanda2011}, we can find their contributions to the 10 LECs of the minimal basis,
\begin{eqnarray}
  E_1&=&-\frac{1}{2} c_{33} - \frac{3}{2} c_{43} + \frac{3}{2} c_{55},\\
  E_2&=&0,\\
  E_3&=&0,\\
  E_4&=& -c_{43}  -\frac{1}{3} c_{47} - \frac{1}{3} c_{51} -2 c_{59} - \frac{2}{3} c_{60}+ \frac{2}{3} c_{61} -\frac{2}{3} c_{62},\\
  E_5&=&0,\\
  E_6&=&  -\frac{1}{3} c_{47} - \frac{1}{3} c_{51}  - \frac{2}{3} c_{60}+ \frac{2}{3} c_{61} +\frac{4}{3} c_{62},\\
  E_7 &=& 24 c_{40} - 24 c_{51} + 8 c_{63} +  8 c_{64},\\
  E_8 &=& 8 c_{40} - 8 c_{51} + 4 c_{63} +  4 c_{64},\\
  E_9 &=&0,\\
  E_{10} &=&c_{37} + c_{40} - c_{47} - c_{51}.
\end{eqnarray}
There are therefore only 6 independent combinations, so that  the large-$N_c$ predictions can be summarized by the following constraints:
\begin{equation} \label{eq:largenc}
  E_2= E_3= E_5= E_9=0.
\end{equation}

The projection onto the $T=1/2$ channel for this restricted interaction, along the same lines of Sec.~\ref{sec:model}, leads to  only five surviving LECs entering the $p-d$ observables. As before, we can effectively impose the constraint
\begin{equation} \label{eq:largenct12}  
  2 E_1 - E_{10}=0
\end{equation}
  and understand that the LECs resulting from the fit are determined only up to an appropriate shift.

\begin{figure}
  \centerline{\includegraphics[width=15cm]{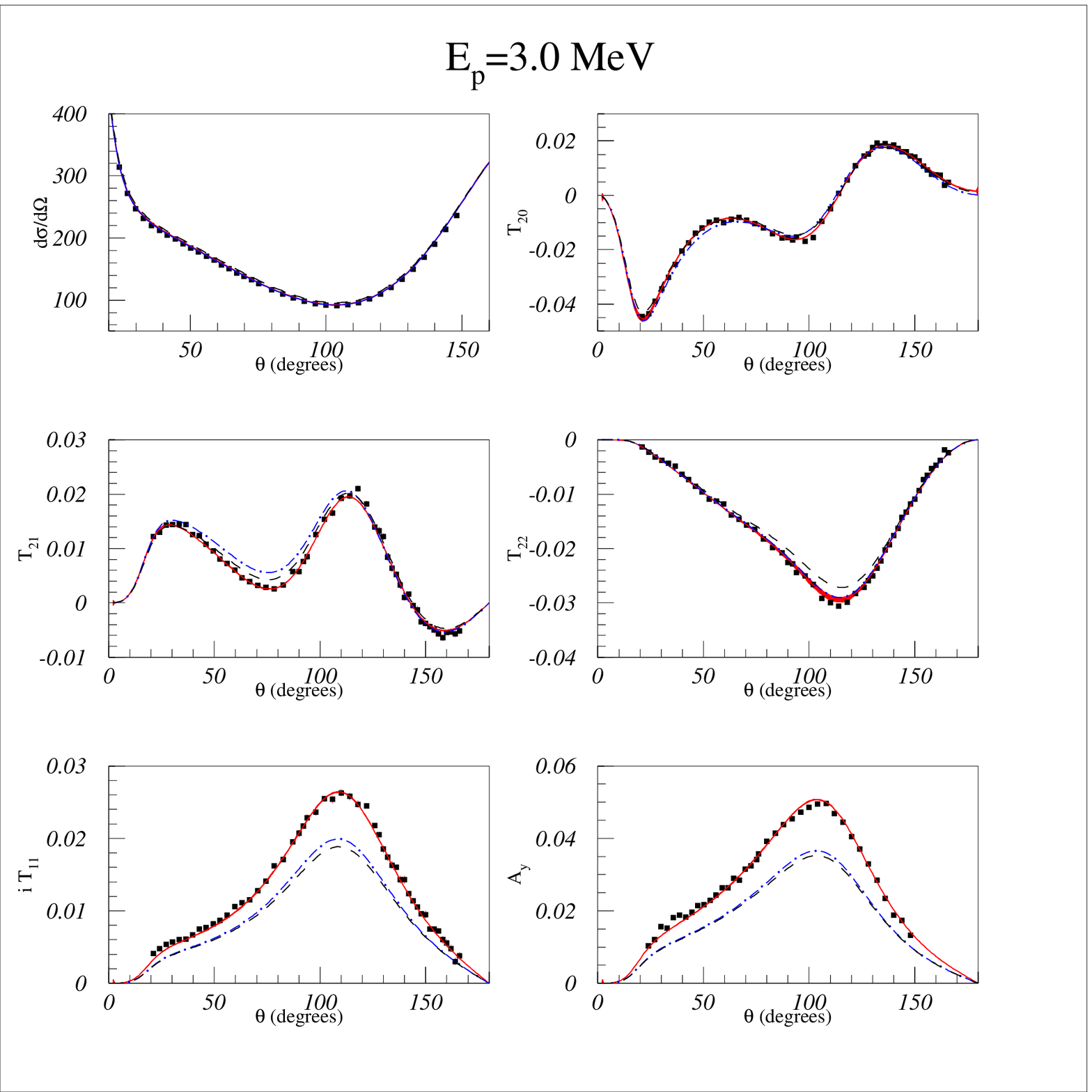}}
  \caption{Same as Fig.~\ref{fig:fit} but for the models which do not include the UIX TNI.\label{fig:fitpl}}
\end{figure}

In order to test these large-$N_c$ predictions, we perform 6-parameter fits to the same experimental data at $E_p=3$~MeV, subjected to the  large-$N_c$ and $T=1/2$ constraints, Eqs.~(\ref{eq:largenc}) and (\ref{eq:largenct12}), with the AV18 $NN$ interaction plus the purely contact TNI of Eq~(\ref{eq:vct}).
We ignore the Urbana IX potential in this case, since it also includes a short-distance component.
For the sake of comparison, we show in Table~\ref{tab:pionless} and Fig.~\ref{fig:fitpl} the fit results obtained ignoring the UIX interaction, which are of the same quality as the ones reported in Sec.~\ref{sec:fit}.
\begin{table}
  \begin{tabular}{|c|c|c|c|c|}
    \hline
$\Lambda $ (MeV)    & 200 &300 & 400 & 500
    \\
    \hline
    $\chi^2$/d.o.f. & 1.6    & 1.7 & 1.7 & 1.7 \\
    $e_0$ & 4.199   & -2.471  &-4.366 & -2.846 \\
    $e_1$ & 2.231   & 2.747  &-3.058 & -2.810 \\
    $e_2$ & -2.846  & -3.084 &3.123 &2.197 \\
    $e_3$ & -1.821  & -3.534  &1.135 &1.247 \\ 
    $e_4$ & 0.753   &3.519  &-0.886 &-1.356 \\
    $e_5$ &-0.547   &0.006  &-0.107 &-0.674 \\
    $e_6$ &0.786    & -0.312  &-0.528 &-0.775 \\
    $e_7$ &7.284    & 3.549  &1.511 &2.287 \\
    $e_8$ &2.260   &0.799  &-0.138 &-0.649 \\
    $e_9$ &-0.271   & -1.709 &0.702 &-0.834 \\
    $e_{10}$ & -0.960& 1.036  & -0.572 &-0.391 \\
%    $B(^3$H) (MeV) & 8.482 & 8.482 & 8.482 &8.482 \\
    \hline
    $a_2$ (fm) & 0.638 & 0.647 & 0.646 &0.647\\ 
    $a_4$ (fm) & 6.32 & 6.32 &6.32 &6.32\\
    \hline
  \end{tabular}
  \caption{Same as Table~\ref{tab:uix} but for the models which do not include  the UIX TNI.\label{tab:pionless}}
\end{table}

\begin{figure}
  \centerline{\includegraphics[width=15cm]{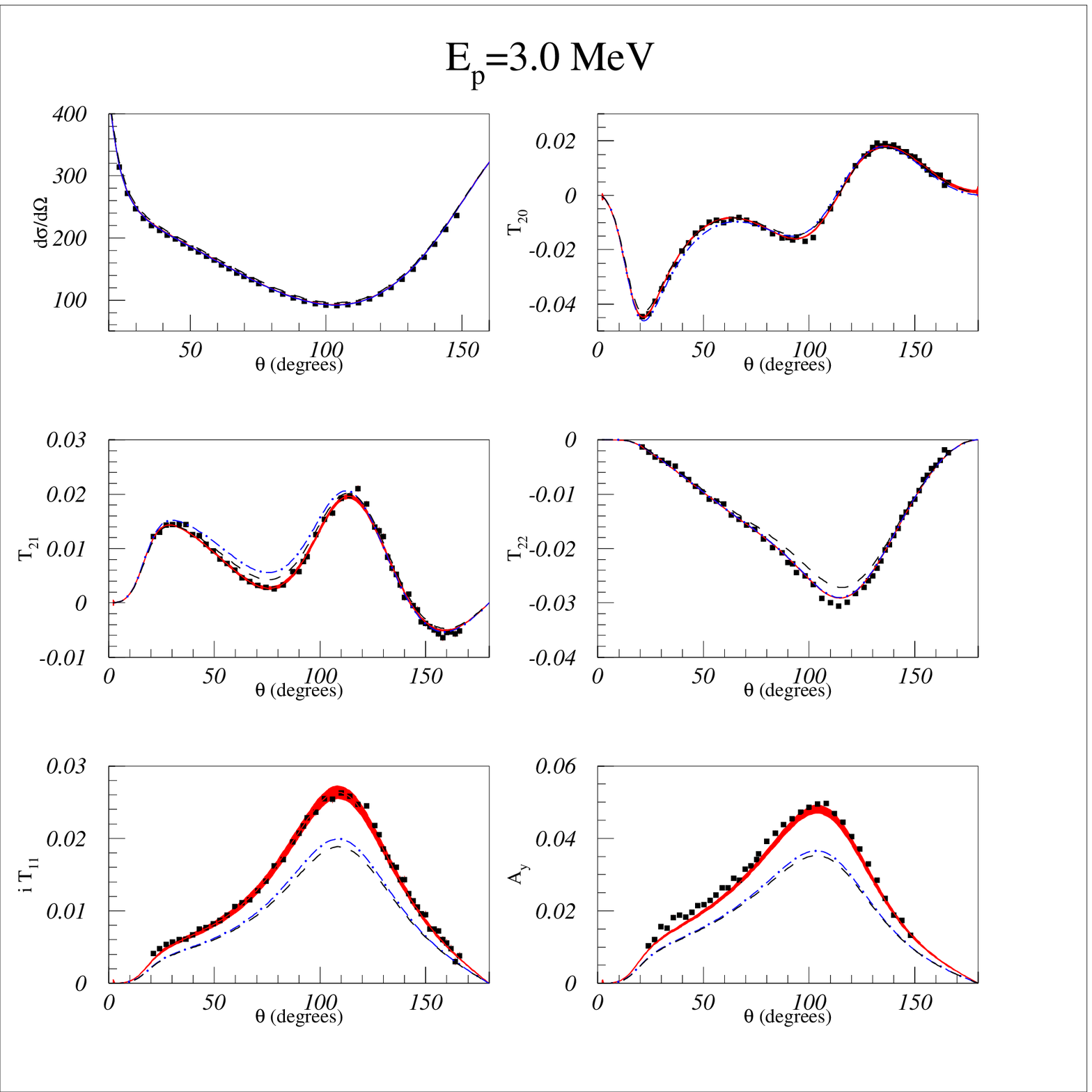}}
  \caption{Fit results in the leading order of the large-$N_c$ limit to a set of cross section and polarization $p-d$ observables at 3~MeV proton energy, for $\Lambda=200-500$~MeV (red bands) as compared to the purely two-body AV18 interaction (dashed, black lines) and to the AV18+UIX two- and three-nucleon interaction (dashed-dotted, blue lines). \label{fig:fitnc}}
\end{figure}

The results corresponding to the leading order of the large-$N_c$ expansion are shown in Table~\ref{tab:largenc} and Fig.~\ref{fig:fitnc}.
\begin{table}
  \begin{tabular}{|c|c|c|c|c|}
    \hline
$\Lambda $ (MeV)    & 200 &300 & 400 & 500
    \\
    \hline
    $\chi^2$/d.o.f. & 2.0 & 1.9 & 2.0 & 2.1 \\
    $e_0$ & -0.309 & -1.766 &0.297 & -0.953 \\
    $e_1$ & 0.328 & 0.400 &0.615 & 0.717 \\
    $e_4$ & -0.360 &0.317 &-0.349 & 0.593 \\
    $e_6$ & -0.060 &0.115 &0.289 &0.297 \\ 
    $e_7$ & 10.093 &15.037 &17.641 &17.862 \\
    $e_8$ &3.559 & 5.371 &5.503 &5.961 \\
    $e_{10}$ & 0.655 &0.800  & 1.230 &1.433 \\
%    $B(^3$H) (MeV) & 8.482 & 8.482 & 8.482 &8.482 \\
    \hline
    $a_2$ (fm) & 0.668 & 0.653 & 0.629 &0.628\\ 
    $a_4$ (fm) & 6.32 & 6.32 &6.32 &6.32\\
    \hline
  \end{tabular}
  \caption{Same as Table~\ref{tab:uix} for the models corresponding to the leading order of the large-$N_c$ expansion.\label{tab:largenc}}
\end{table}
By inspection of the table we can conclude that  reasonable fits can be obtained in this limit, although at the cost of unnatural values for the spin-orbit LECs $E_7$. The increasing $\chi^2$/d.o.f. for higher  values of $\Lambda$ might be the consequence of the absence of longer-range components in the TNI from pion-exchange contributions, which would also be leading in the large-$N_c$ counting.

\subsection{Relativistic counting} \label{sec:rel}
A different kind of hierarchy among subleading contact operators can be deduced in the framework of the recently proposed relativistic counting for the $NN$ contact operators \cite{Ren2018}. In this approach, one retains, in the leading-order Lagrangian,  all relativistically invariant 4-nucleon operators involving no spacetime derivatives,
\begin{eqnarray}
  {\cal L}_{NN}^{(0)} &=& \frac{1}{2}\left[C_S (\bar \psi \psi) (\bar \psi \psi) + C_P (\bar \psi \gamma_5 \psi ) ( \bar \psi \gamma_5 \psi) + C_V (\bar \psi \gamma^\mu \psi ) ( \bar \psi \gamma_\mu \psi) \right. \nonumber \\
    &&\left. + C_{A} ( \bar \psi \gamma^\mu \gamma_5 \psi) ( \bar \psi \gamma_\mu \gamma_5 \psi) + C_T (\bar \psi \sigma^{\mu\nu} \psi ) ( \bar \psi \sigma_{\mu\nu} \psi) \right],
\end{eqnarray}
where $\psi$ collects the Dirac spinor nucleon fields. In contrast to common practice, one does not expand around the static nucleon limit, which would amount to collapsing the 5 LECs onto 2 independent combinations, Eq~(\ref{eq:nncontact}), which parametrize the central and spin-spin short-range potential. Instead, all 5 LECs are considered on an equal footing,  generating further spin operators, among which the spin-orbit term. This procedure yields a much faster convergence of the low-energy expansion, since at each order there are more adjustable parameters.
We can apply the same procedure to the three-nucleon case by writing all possible relativistically invariant 6-nucleon operators, symmetric under isospin, charge-conjugation (${\cal C}$), parity (${\cal P}$) and time reversal (${\cal T}$) transformations.
The transformation properties of the different space-time and isospin structure inside fermion bilinears under the discrete symmetries are displayed in Table~\ref{tab:cpt}. 
\begin{table}[bth]
\begin{center}
\begin{tabular}{l|c|c|c|c|c|c|c}
\hline
\hline
& 1& $\gamma_5$ & $ \gamma_\mu$ & $\gamma_\mu \gamma_5$ & $\sigma_{\mu \nu}$
& $\tau^a$ &$ \epsilon_{\mu \nu \rho \sigma}$  \\
\hline
${\cal P}$ & + & -- & + & -- & + &  + & --  \\
\hline
${\cal C}$ & + & + & -- & + & -- & $(-1)^{a+1}$ & + \\
\hline
H.c.\ & + & -- & + & + & + & + & + \\
\hline
\hline
\end{tabular}
\end{center}
\caption{Transformation properties of the fermion bilinears
with the different elements of the Clifford and flavour algebra, and Levi-Civita tensor under parity (${\cal P}$), charge conjugation (${\cal C}$), and Hermitian conjugation (H.c.).}
 \label{tab:cpt}
\end{table}
Based on these properties we can form a set of 25 different
 operators, displayed in Table~\ref{tab:relop}.
\begin{table}
 \begin{tabular}{|l|l|}
\hline
$O^{\mathrm{rel}}_{1,2}$ &$(\bar\psi \psi)_1 (\bar\psi \psi)_2 (\bar\psi \psi)_3  \otimes [ {\bf 1}, {\bm \tau}_1\cdot {\bm \tau}_2 ]$\\
$O^{\mathrm{rel}}_{3,4,5}$& $(\bar\psi  \psi)_1 (\bar\psi \gamma_5 \psi)_2 (\bar\psi \gamma_5 \psi)_3 \otimes [ {\bf 1}, {\bm \tau}_2\cdot {\bm \tau}_3, {\bm \tau}_1\cdot ( {\bm \tau}_2 +{\bm \tau}_3)]$ \\
$O^{\mathrm{rel}}_{6,7,8}$ &$(\bar\psi  \psi)_1 (\bar\psi \gamma^\mu \psi)_2 (\bar\psi \gamma_\mu \psi)_3 \otimes[ {\bf 1}, {\bm \tau}_2\cdot {\bm \tau}_3, {\bm \tau}_1\cdot ( {\bm \tau}_2 +{\bm \tau}_3)]$ \\
$O^{\mathrm{rel}}_{9,10,11}$& $(\bar\psi  \psi)_1 (\bar\psi \gamma^\mu\gamma_5 \psi)_2 (\bar\psi \gamma_\mu \gamma_5 \psi)_3 \otimes [ {\bf 1}, {\bm \tau}_2\cdot {\bm \tau}_3, {\bm \tau}_1\cdot ( {\bm \tau}_2 +{\bm \tau}_3)]$\\
$O^{\mathrm{rel}}_{12,13,14}$&  $(\bar\psi  \psi)_1 (\bar\psi \sigma^{\mu\nu} \psi)_2 (\bar\psi \sigma_{\mu\nu} \psi)_3 \otimes[ {\bf 1}, {\bm \tau}_2\cdot {\bm \tau}_3, {\bm \tau}_1\cdot ( {\bm \tau}_2 +{\bm \tau}_3)]$\\
$O^{\mathrm{rel}}_{15}$ & $(\bar\psi  \gamma_5 \psi)_1 (\bar\psi \gamma^\mu \psi)_2 (\bar\psi \gamma_\mu \gamma_5 \psi)_3 \otimes [{\bm \tau}_1 \cdot {\bm \tau}_2 \times {\bm \tau}_3]$ \\
$O^{\mathrm{rel}}_{16}$& $(\bar\psi  \sigma^{\mu\nu} \psi)_1 (\bar\psi \gamma_\mu \psi)_2 (\bar\psi \gamma_\nu  \psi)_3 \otimes [{\bm \tau}_1 \cdot {\bm \tau}_2 \times {\bm \tau}_3]$
\\
$O^{\mathrm{rel}}_{17}$& $(\bar\psi  \sigma^{\mu\nu} \psi)_1 (\bar\psi \gamma_\mu \gamma_5 \psi)_2 (\bar\psi \gamma_\nu  \gamma_5 \psi)_3 \otimes [{\bm \tau}_1 \cdot {\bm \tau}_2 \times {\bm \tau}_3]$
\\
$O^{\mathrm{rel}}_{18}$& $(\bar\psi  \sigma^{\mu\nu} \psi)_1 (\bar\psi \sigma_{\mu \alpha} \psi)_2 (\bar\psi \sigma^\alpha_\nu \psi)_3 \otimes [{\bm \tau}_1 \cdot {\bm \tau}_2 \times {\bm \tau}_3]$
\\
$O^{\mathrm{rel}}_{19,20,21}$ & $ (\bar\psi  \gamma_5 \psi)_1 (\bar\psi \sigma^{\mu\nu} \psi)_2 (\bar\psi \sigma_{\mu\nu} \gamma_5 \psi)_3 \otimes [ {\bf 1}, {\bm \tau}_2\cdot {\bm \tau}_3, {\bm \tau}_1\cdot ( {\bm \tau}_2 +{\bm \tau}_3)]$
\\
$O^{\mathrm{rel}}_{22,23,24,25}$ & $ (\bar\psi  \gamma_\mu \psi)_1 (\bar\psi \gamma_{\nu} \gamma_5 \psi)_2 (\bar\psi \sigma^{\mu\nu} \gamma_5 \psi)_3 \otimes [ {\bf 1}, {\bm \tau}_2\cdot {\bm \tau}_3, {\bm \tau}_1\cdot ( {\bm \tau}_2 +{\bm \tau}_3), {\bm \tau}_1\cdot ({\bm \tau}_2 - {\bm \tau}_3)]$\\
\hline
 \end{tabular}
 \caption{\label{tab:relop}A complete, but non-minimal, set of Lorentz, isospin, ${\cal C}$, ${\cal P}$, ${\cal T}$-invariant $3N$ contact operators involving no spacetime derivatives of fields.}
\end{table}

Simultaneous rearrangements of Dirac and flavour indices between identical nucleon fields lead to Fierz identities, as detailed in Appendix~\ref{app:fierz}. As a result, the leading relativistic $3N$ contact Lagrangian is written in terms of 5 independent operators,
\begin{eqnarray} \label{eq:relp0}
  {\cal L}_{3N}^{(0)} &=& -\left[E_S (\bar \psi \psi) (\bar \psi \psi) + E_P (\bar \psi \gamma_5 \psi ) ( \bar \psi \gamma_5 \psi) + E_V (\bar \psi \gamma^\mu \psi ) ( \bar \psi \gamma_\mu \psi) \right. \nonumber \\
    &&\left. + E_{A} ( \bar \psi \gamma^\mu \gamma_5 \psi) ( \bar \psi \gamma_\mu \gamma_5 \psi) + E_T (\bar \psi \sigma^{\mu\nu} \psi ) ( \bar \psi \sigma_{\mu\nu} \psi) \right] (\bar \psi \psi), \nonumber \\
  &\equiv& -\sum_X E_X O_X, \quad X=S,P,V,A,T.
\end{eqnarray}
The non-relativistic expansion of the nucleon fields,
\begin{equation}
  \psi(x) =  \left( \begin{array}{c}  1 + \frac{\nabla^2}{8 m^2} \\
  - \frac{i}{2m}   {\bm \sigma} \cdot {\bm \nabla} \end{array} \right)
 N(x)
  + {\cal O}(Q^3)  \ .\label{eq:psiN}
\end{equation}
allows to express the 5 operators $O_X$ in terms of the subleading operators defined in Ref.~\cite{Girlanda2011} as
\begin{eqnarray}
  O_S &=& (N^\dagger N)^3 + \frac{3}{8 m^2} \left[ o_{127} -2  o_{75}\right],\\
  O_P &=& -\frac{1}{4 m^2} o_{36}, \\
  O_V &=& (N^\dagger N)^3 -\frac{1}{8 m^2} \left[ 4 o_{33} -4 o_{75} - 4 o_{79} -2 o_1 +2 o_{42} -2 o_{39} -o_{127} +2 o_{75} \right],\\
  O_A &=& (N^\dagger N)^3 -\frac{1}{8 m^2} \left[ 2 o_4 + 2 o_{130} -2 o_{36} -2 o_{45} -2 o_{137} - 2 o_{79} - 2 o_{83} + o_{134} \right. \nonumber \\
    && \left. -2 o_{115}\right],\\
  O_T &=& -2 (N^\dagger N)^3 -\frac{1}{4 m^2} \left[ 2 o_{33} - 2 o_{10} + 2 o_{7} - 4 o_{75} +2 o_{42} + 2 o_{53} - 2 o_{36} - 2 o_{45} - 2 o_{137} \right.\nonumber \\
    &&\left. - 2 o_{79} - 2 o_{83} - o_{134} + 2 o_{115} \right],
\end{eqnarray}
where the relation
\begin{equation}
  (N^\dagger N)^3 = - (N^\dagger {\bm \sigma} N) \cdot (N^\dagger {\bm \sigma } N) (N^\dagger N) 
\end{equation}
has been used, also a consequence of Fierz identities.
It is possible to express the above operators in the minimal basis \cite{Girlanda2011}, obtaining
\begin{eqnarray}
  O_S &=& O_0 + \frac{1}{m^2} \left[ -\frac{3}{16} O_1 - \frac{1}{8} O_2 - \frac{1}{8} O_3 - \frac{1}{6} O_4 + \frac{1}{16} O_5 + \frac{1}{48} O_6 - \frac{5}{4} O_7 - \frac{1}{4} O_8 \right],\\
  O_P &=& -\frac{1}{4 m^2} O_9, \\
  O_V &=& O_0 + \frac{1}{m^2} \left[ \frac{3}{16} O_1 - \frac{1}{8} O_4 - \frac{3}{16} O_5 - \frac{1}{16} O_6 - \frac{1}{4} O_7 -\frac{1}{4} O_8 - \frac{1}{2} O_9 - \frac{1}{4} O_{10} \right],\\
  O_A &=& O_0 + \frac{1}{m^2} \left[ \frac{3}{16} O_1 - \frac{1}{4} O_2 + \frac{1}{4} O_3 - \frac{1}{24} O_4 + \frac{1}{16} O_5 + \frac{1}{48} O_6 + \frac{11}{4} O_7 +\frac{3}{4} O_8 +\frac{1}{2} O_9 \right. \nonumber \\
    && \left. + \frac{1}{4} O_{10} \right],\\
  O_T &=& -2 O_0 + \frac{1}{m^2} \left[ \frac{3}{8} O_1 + \frac{3}{4} O_2 - \frac{1}{4} O_3 + \frac{1}{6} O_4 - \frac{5}{8} O_5 - \frac{5}{24} O_6 - \frac{7}{2} O_7 -\frac{3}{2} O_8 \right. \nonumber\\
    &&\left. -\frac{1}{2} O_9  \right],
\end{eqnarray}
where the operator $O_0$ contains the relativistic ``drift corrections'' \cite{Girlanda2010},
\begin{eqnarray}
  O_0&=&(N^\dagger N)^3 + \frac{1}{16 m^2} \left[ 3 (N^\dagger \nrd N)\cdot (N^\dagger \nrd N) (N^\dagger N) - (N^\dagger \nrd \tau^a N)\cdot (N^\dagger \nrd \tau^a N) (N^\dagger N) \right. \nonumber \\
    &&\left. - 3 (N^\dagger \nrd \cdot {\bm \sigma} N) (N^\dagger \nrd \cdot {\bm \sigma} N) (N^\dagger N) - (N^\dagger \nrd \cdot {\bm \sigma} \tau^a N) (N^\dagger \nrd \cdot {\bm \sigma} \tau^a N) (N^\dagger N) \right].
\end{eqnarray}
Neglecting the latter we have
\begin{align} 
  E_0 &=E_S + E_V + E_A - 2 E_T, &&  \label{eq:rele0}\\ 
  m^2 E_1 &= - \frac{3}{16} E_S + \frac{3}{16} E_V + \frac{3}{16} E_A +\frac{3}{8} E_T &=&
  -\frac{3}{16} E_0 +\frac{3}{8} E_V +\frac{3}{8} E_A,   \label{eq:rele1}\\
 m^2 E_2 &=  - \frac{1}{8} E_S - \frac{1}{4} E_A +\frac{3}{4} E_T &=&
-\frac{1}{8} E_0 +\frac{1}{8} E_V -\frac{1}{8} E_A +\frac{1}{2} E_T,   \label{eq:rele2}\\
  m^2 E_3 &= - \frac{1}{8} E_S + \frac{1}{4} E_A -\frac{1}{4} E_T &=&
-\frac{1}{8} E_0 +\frac{1}{8} E_V +\frac{3}{8} E_A -\frac{1}{2} E_T,   \label{eq:rele3}\\
m^2 E_4 &= - \frac{1}{6} E_S - \frac{1}{8} E_V - \frac{1}{24} E_A +\frac{1}{6} E_T &=&
-\frac{1}{6} E_0 +\frac{1}{24} E_V +\frac{1}{8} E_A -\frac{1}{6} E_T,   \label{eq:rele4}\\
m^2 E_5 &=  \frac{1}{16} E_S - \frac{3}{16} E_V + \frac{1}{16} E_A -\frac{5}{8} E_T &=&
\frac{1}{16} E_0 -\frac{1}{4} E_V  -\frac{1}{2} E_T,  \label{eq:rele5}\\
m^2 E_6 &=  \frac{1}{48} E_S - \frac{1}{16} E_V + \frac{1}{48} E_A -\frac{5}{24} E_T &=&
\frac{1}{48} E_0 -\frac{1}{12} E_V  -\frac{1}{6} E_T,   \label{eq:rele6}\\
m^2 E_7 &=  -\frac{5}{4} E_S - \frac{1}{4} E_V + \frac{11}{4} E_A -\frac{7}{2} E_T &=&
-\frac{5}{4} E_0 + E_V +4 E_A -6 E_T,  \label{eq:rele7}\\
m^2 E_8 &= -\frac{1}{4} E_S - \frac{1}{4} E_V + \frac{3}{4} E_A -\frac{3}{2} E_T &=&
-\frac{1}{4} E_0 + E_A -2 E_T,   \label{eq:rele8}\\
m^2 E_9 &=  -\frac{1}{4} E_P - \frac{1}{2} E_V + \frac{1}{2} E_A -\frac{1}{2} E_T ,&&   \label{eq:rele9}\\
  m^2 E_{10} &=  - \frac{1}{4} E_V + \frac{1}{4} E_A  ,& & \label{eq:rele10}
  \end{align}
with only 4 independent combinations of the subleading $3N$ contact LECs. We notice in particular that the numerical coefficients entering in the expression of $E_7$ are larger by one order of magnitude compared to the other LECs. This might be at the origin of the phenomenological prominence of the spin-orbit interaction encoded in $E_7$, already proposed in Ref.~\cite{Kievsky1999}.

In  terms of the operators defined in Table~\ref{tab:relop} the isospin projection reads
\begin{align}
  (O_1^{\mathrm{rel}})_{1/2}&= \frac{1}{2} O_1^{\mathrm{rel}} - \frac{1}{2} O_2^{\mathrm{rel}} &=& \frac{5}{4} O_S + \frac{1}{4} O_P + \frac{1}{4} O_V - \frac{1}{4} O_A + \frac{1}{8} O_T,\\
  (O_3^{\mathrm{rel}})_{1/2}&= \frac{1}{2} O_3^{\mathrm{rel}} - \frac{1}{6} O_4^{\mathrm{rel}}- \frac{1}{6} O_5^{\mathrm{rel}} &=& -\frac{1}{12} O_S + \frac{11}{12} O_P - \frac{1}{12} O_V + \frac{1}{12} O_A - \frac{1}{24} O_T,\\
  (O_6^{\mathrm{rel}})_{1/2}&= \frac{1}{2} O_6^{\mathrm{rel}} - \frac{1}{6} O_7^{\mathrm{rel}}- \frac{1}{6} O_8^{\mathrm{rel}} &=& -\frac{1}{3} O_S - \frac{1}{3} O_P + \frac{2}{3} O_V + \frac{1}{3} O_A - \frac{1}{6} O_T,\\
  (O_9^{\mathrm{rel}})_{1/2}&= \frac{1}{2} O_9^{\mathrm{rel}} - \frac{1}{6} O_{10}^{\mathrm{rel}}- \frac{1}{6} O_{11}^{\mathrm{rel}} &=& \frac{1}{3} O_S + \frac{1}{3} O_P + \frac{1}{3} O_V + \frac{2}{3} O_A +\frac{1}{6} O_T,\\
  (O_{12}^{\mathrm{rel}})_{1/2}&= \frac{1}{2} O_{12}^{\mathrm{rel}} - \frac{1}{6} O_{13}^{\mathrm{rel}}- \frac{1}{6} O_{14}^{\mathrm{rel}} &=& - O_S - O_P - O_V +  O_A +\frac{1}{2} O_T,
\end{align}
whence one conclude that there is only one purely $T=3/2$ operator,
\begin{equation}
  O^{\mathrm{rel}}_{3/2}=O_S + O_P + O_V - O_A + \frac{1}{2} O_T,
\end{equation}
and 4 purely $T=1/2$ combinations, e.g.,
\begin{equation}
  (O_S + 3 O_P), \quad (O_V -4 O_P), \quad (O_V+ O_A), \quad (3 O_V - O_T).
\end{equation}
As before, in fitting to $p-d$ observables, one can impose the constraint
\begin{equation} \label{eq:t12rel}
  3 E_S - E_P - 4 E_V + 4 E_A -12 E_T=0,
  \end{equation}
with the understanding that the determination of the LECs is only valid up to a shift involving $h_{3/2}$.

\begin{figure}
  \centerline{\includegraphics[width=15cm]{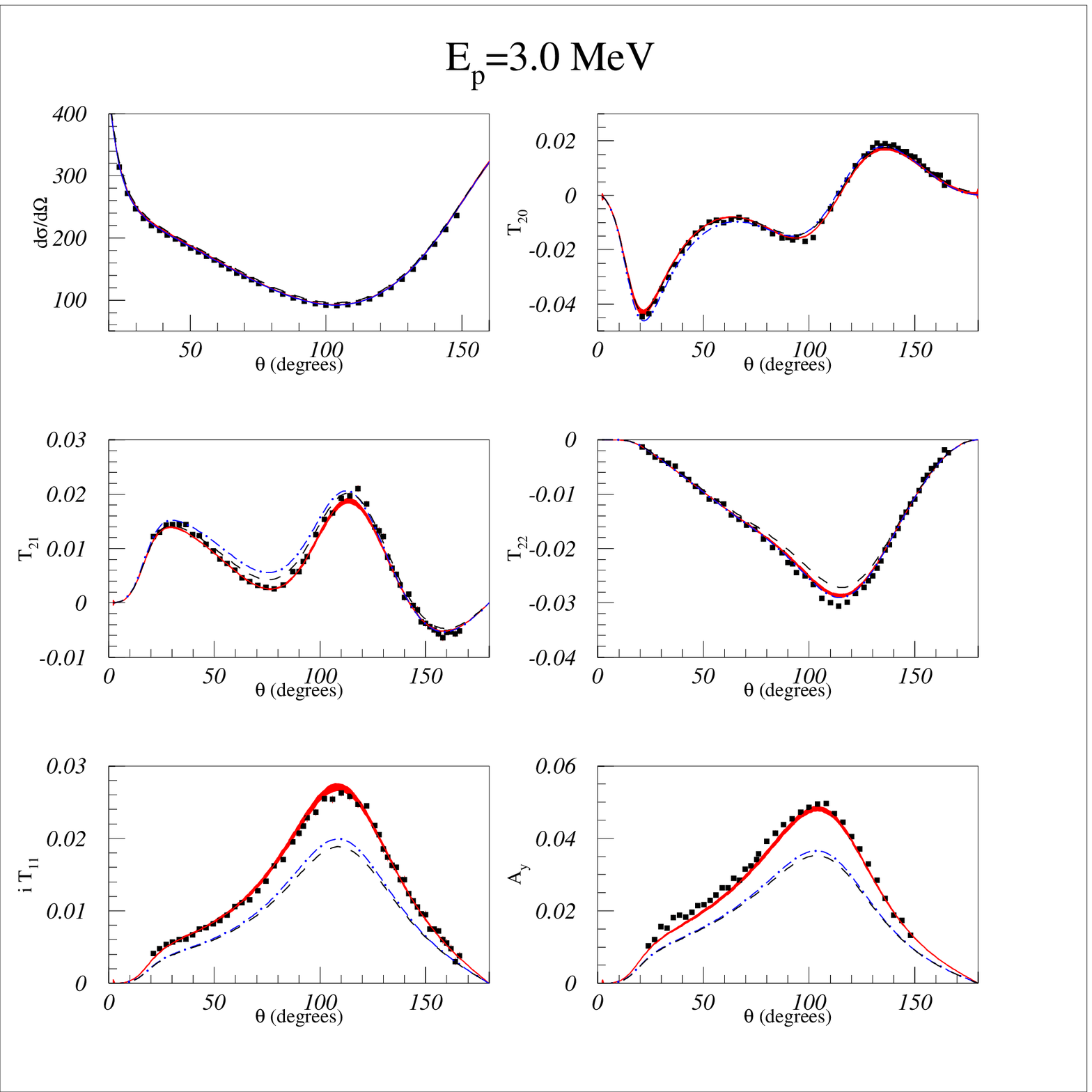}}
  \caption{Fit results in the leading order of the relativistic counting to a set of cross section and polarization $p-d$ observables at 3~MeV proton energy, for $\Lambda=200-500$~MeV (red bands) as compared to the purely two-body AV18 interaction (dashed, black lines) and to the AV18+UIX two- and three-nucleon interaction (dashed-dotted, blue lines). \label{fig:fitrel}}
\end{figure}

In order to test the effectiveness of the relativistic counting, we also fit the constants $E_X$, $X=S,P,V,A,T$ subjected to the $T=1/2$ constraint. As in the case of the large-$N_c$ limit we test this description in the framework of a purely pionless TNI. Specifically we consider a nuclear Hamiltonian consisting of the AV18 $NN$ interaction and the leading and subleading $3N$ contact terms implied by the Lagrangian~(\ref{eq:relp0}). The latter can be expressed in the usual basis expressing the LECs $E_i$'s in terms of the $E_X$'s using the relations (\ref{eq:rele0})-(\ref{eq:rele10}), with the $T=1/2$ constraint expressed in Eq.~(\ref{eq:t12rel}).
The fitted parameters are the adimensional $e_0$, $e_V$, $e_A$ and $e_T$, having defined the adimensional $e_X$ as
\begin{equation}
  E_X=\frac{e_X}{F_\pi^4\Lambda},\quad X=V,A,T.
\end{equation}
The results are displayed in Table~\ref{tab:rel}. Compared to the unconstrained 10-parameter fit, the $\chi^2/$d.o.f. is slightly increased, but a reasonable description is obtained for all adopted values of $\Lambda$. This may be considered as compatible with the leading-order character of the interaction, and gives support to the relativistic counting in the three-nucleon sector.
\begin{table}
  \begin{tabular}{|c|c|c|c|c|}
    \hline
$\Lambda$ (MeV)    & 200 & 300 & 400 & 500
    \\
    \hline
    $\chi^2$/d.o.f. & 2.2 & 2.2 & 2.3 & 2.1 \\
    $e_0$ & -1.169 & -0.565 &0.130 & 1.506 \\
    $e_1$ & 0.202 & 0.245   &0.376 & 1.077 \\
    $e_2$ & 0.074& 0.031 &0.028 &-0.193 \\
    $e_3$ & 0.067 &0.139 &0.219 &0.857  \\
    $e_4$ & 0.029 &0.054 &0.070 & 0.232 \\
    $e_5$ & -0.135 &-0.106 &-0.157 &-0.219 \\
    $e_6$ & -0.045 &-0.035 &-0.052 &-0.073 \\ 
    $e_7$ & 0.533 &1.343 &2.127 &8.850 \\
    $e_8$ &-0.002 &0.230 &0.375 &1.993 \\
    $e_9$ & 0.974 &0.606 &2.274 &4.552 \\
    $e_{10}$ & -0.415 &-0.046 &  -1.468 &-2.837 \\
%    $B(^3$H) (MeV) & 8.482 & 8.482 & 8.482 &8.482 \\
\hline
    $a_2$ (fm) & 0.707 & 0.619 & 0.639 &0.646 \\ 
    $a_4$ (fm) & 6.32 & 6.32 &6.32 &6.32\\
    \hline
  \end{tabular}
\caption{Same of Table~\ref{tab:uix} for the model restricted at the leading-order of the relativistic counting.\label{tab:rel}}
\end{table}
We show in Fig.~\ref{fig:fitrel} the corresponding description of the $p-d$ scattering observables.

\section{Conclusions} \label{sec:concl}

The aims of this paper are twofold. In the first part we discuss the possibility of determining the
subleading contact three-body interaction from a fit procedure of selected binding energies and scattering data.
In particular the fit includes the triton binding energy, the doublet and quartet $n-d$ scattering lengths
and several $p-d$ scattering data. Being this the first attempt to incorporate systematically scattering data 
in the determination of the TNI, we limit the fit of the $p-d$ data to a single energy, $E_p=3$~MeV, at which
around 300 data point exist. However we
plan to extend the fit procedure to a simultaneous inclusion of data at several energies, 
below and above the deuteron breakup threshold. Regarding the results of the present analysis,
they were satisfactory: the $\chi^2$/d.o.f obtained after the fit was below 2 similar to the
values obtained in the fit of the $NN$ potential in the two-nucleon sector. Moreover they show a very small cutoff dependence. In this way
we have shown that the subleading contact terms of the TNI provide enough flexibility to fit satisfactorily  
low-energy elastic $N-d$ scattering observables, thus solving the long-standing  discrepancy in some polarization observables. After decades of strong
efforts to describe the $NN$ database with values of $\chi^2$/d.o.f close to 1, similar values were impossible to achieve
in the three-nucleon sector even with the inclusion of very sophisticated TNI as the UIX type or from ChEFT up to N3LO.
The angular-spin-isospin dependence of the TNI seems to be much more complicated than the forms used up to
now. Here we have shown that this dependence  can be opportunely collected in the sum of the 10 terms given by the contact N4LO interaction.

To perform the fit the TNI contact interaction was summed to the AV18+UIX, widely used in the description
of nuclear states.
The fit at $E_p=3$~MeV  determines the 10 combinations of LECs, $e_i,\; i=0,\ldots,10$ (one LEC is fixed by the condition $h_{3/2}=0$) relevant to the $T=1/2$ channel. 
To evaluate the capabality of the complete potential to describe other data, we explored the low-energy region,
$E_p<3$~MeV, in which
several observables have been measured. We have observed an overall good agreement with a satisfactory 
description of the vector analyzing powers down to very low energies.

The relative importance of the  10 subleading terms has been discussed in the second part of the paper.
In fact, substantial improvement in the description of the same $p-d$ observables is also provided by 
simplified versions of this interaction, given by the 
leading order of  the recently proposed relativistic counting \cite{Ren2018} or of the large-$N_c$ expansion \cite{thooft1974,Kaplan1997,Kaplan1996,Phillips2013}. Eventhough the naturality 
and the cutoff dependence of the involved LECs cannot be properly addressed, since we are using purely phenomenological 
models as the bulk of the $NN$ and $3N$ interactions, the results suggest natural values of the LECs and a very mild cutoff 
dependence of the theoretical description in the range $\Lambda=200-500$~MeV.
It will be interesting to study this interaction in conjunction with chiral $NN$ and $3N$ potentials or with purely contact 
nuclear interactions as implied in pionless EFT.
Further investigation is also needed in order to test these models at higher energies and in larger systems. 
Studies along these lines are in progress.

\begin{acknowledgments}
  The Authors  would like to thank Rocco Schiavilla for very useful discussions.
  \end{acknowledgments}

\appendix
\section{Fierz-like identities}
\label{app:fierz}
In this appendix we detail the Fierz-type relations used to reduce the relativistic operators in Sec.~\ref{sec:rel}.
The rearrangement of indices, indicated by round and square brackets for the two involved bilinears, concern  Dirac indices,  as e.g.
\begin{equation}
  ({\bf 1} ) [{\bf 1}] = \frac{1}{4} \left\{ ({\bf 1} ] [ {\bf 1} ) +  ( \gamma_5 ] [ \gamma_5 ) +  ( \gamma^\mu ] [ \gamma_\mu ) - (\gamma^\mu \gamma_5 ] [ \gamma_\mu \gamma_5 ) + \frac{1}{2} ( \sigma^{\mu \nu} ] [ \sigma_{\mu \nu} )\right\},
\end{equation}
and isospin indices, as
\begin{equation}
( {\bf 1} ) [ {\bf 1} ] = \frac{1}{2} ({\bf 1} ] [ {\bf 1}) + \frac{1}{2} (\vec{\tau} ] \cdot [{\vec{\tau}} ).
\end{equation}
Imposing the antisymmetry under exchange of identical fermion fields leads to relations (we omit in this appendix the superscript ``rel'' for the sake of clarity) like
\begin{eqnarray}
  O_1&=&-\frac{1}{8} \left(O_1 + O_3 + O_6 - O_9 + \frac{1}{2} O_{12} + O_2 + O_4 + O_7 - O_{10} + \frac{1}{2} O_{13} \right),\\
  O_2&=&-\frac{1}{8} \left[ 3 \left( O_1 + O_3 + O_6 - O_9 + \frac{1}{2} O_{12}\right) - O_2 - O_4 - O_7 + O_{10} - \frac{1}{2} O_{13} \right].
\end{eqnarray}
Similar relations can be obtained by using further identities, which can be obtained using the completeness of the Dirac bilinears,
{\allowdisplaybreaks
\begin{eqnarray}
  ({\bf 1}) [\gamma_5] &=& \frac{1}{4} \biggl\{ (\gamma_5 ] [ {\bf 1} ) +  ({\bf 1}][\gamma_5) +  (\gamma^\mu \gamma_5 ][\gamma_\mu ) -  (\gamma^\mu ][\gamma_\mu \gamma_5) +  2(\sigma^{\mu\nu}][\sigma_{\mu\nu}\gamma_5) \biggr\},\\
      ({\bf 1} ) [\gamma^\mu ] &=& \frac{1}{4} \biggl\{ (\gamma^\mu ][{\bf 1}) + ({\bf 1}][\gamma^\mu ) - (\gamma^\mu\gamma_5][\gamma_5) + (\gamma_5 ][\gamma^\mu \gamma_5) - i (\gamma_\nu][\sigma^{\mu\nu}) + i (\sigma^{\mu\nu}][\gamma_\nu) \nonumber \\
      &&
      + i (\gamma_\nu \gamma_5][\sigma^{\mu\nu}\gamma_5) + i (\sigma^{\mu\nu}\gamma_5][\gamma_\nu\gamma_5) \biggr\},\\
      (\gamma_5)[\gamma^\mu \gamma_5] &=& \frac{1}{4} \biggl\{(\gamma_5][\gamma^\mu\gamma_5) + (\gamma^\mu \gamma_5][\gamma_5) +({\bf 1}][\gamma^\mu) - (\gamma^\mu][{\bf 1}) + i (\sigma^{\mu\nu}][\gamma_\nu) +i (\gamma_\nu][\sigma^{\mu\nu}) \nonumber \\
      &&+ i (\sigma^{\mu\nu}\gamma_5][\gamma_\nu\gamma_5) - i(\gamma_\nu\gamma_5][\sigma^{\mu\nu}\gamma_5) \biggr\},\\ 
      ({\bf 1})[\gamma^\mu \gamma_5] &=&
      \frac{1}{4} \biggl\{ ({\bf 1}][\gamma^\mu \gamma_5 ) + (\gamma^\mu \gamma_5][{\bf 1}) - (\gamma^\mu][\gamma_5) + (\gamma_5][\gamma^\mu) + i (\gamma_\nu][\sigma^{\mu\nu}\gamma_5)  \nonumber \\
      && + i  (\sigma^{\mu\nu}\gamma_5][\gamma_\nu) -i (\gamma_\nu\gamma_5][\sigma^{\mu\nu}) + i  (\sigma^{\mu\nu}][\gamma_\nu\gamma_5) \biggr\},\\
      ({\bf 1})[\sigma^{\mu\nu}] &=& \frac{1}{4} \biggl\{({\bf 1}][\sigma^{\mu\nu}) + (\sigma^{\mu\nu}][{\bf 1}) + (\gamma_5][\sigma^{\mu\nu}\gamma_5)+ ( \sigma^{\mu\nu}\gamma_5 ][\gamma_5) + i (\sigma^{\mu\alpha}][\sigma_\alpha^{\,\,\,\nu})  \nonumber \\
      && -(\sigma^{\nu\alpha}][\sigma_\alpha^{\,\,\,\mu}) -\epsilon^{\mu\nu\alpha\beta} (\gamma_\alpha][\gamma_\beta \gamma_5) + \epsilon^{\mu\nu\alpha\beta} (\gamma_\alpha \gamma_5][\gamma_\beta ) \biggr\},\\
      (\gamma_5)[\sigma^{\mu\nu}\gamma_5] &=& \frac{1}{4} \biggl\{ i (\gamma^\mu][\gamma^\nu) - i (\gamma^\nu][\gamma^\mu) - i (\gamma^\mu \gamma_5][\gamma^\nu \gamma_5) + i (\gamma^\nu \gamma_5][\gamma^\mu \gamma_5) + ({\bf 1}][\sigma^{\mu\nu})  \nonumber \\
      && + (\sigma^{\mu\nu}][{\bf 1}) + (\gamma_5][\sigma^{\mu\nu}\gamma_5) + (\sigma^{\mu\nu}\gamma_5][\gamma_5) + i (\sigma^{\mu\alpha}][\sigma_{\alpha}^{\,\,\,\nu})- i(\sigma^{\nu\alpha}][\sigma_{\alpha}^{\,\,\,\mu})\nonumber \\
      &&+ \epsilon^{\mu\nu\alpha\beta}(\gamma_\alpha][\gamma_\beta \gamma_5) - \epsilon^{\mu\nu\alpha\beta}(\gamma_\alpha\gamma_5][\gamma_\beta) \biggr\},\\
      (\gamma_5)[\gamma^\mu] &=& \frac{1}{4} \biggl\{ (\gamma_5][\gamma^\mu) + (\gamma^{\mu}][\gamma_5) + ({\bf 1}][\gamma^\mu \gamma_5) -(\gamma^\mu \gamma_5][{\bf 1}) -i (\gamma_\nu][\sigma^{\mu\nu}\gamma_5)  \nonumber \\
      &&+ i (\sigma^{\mu\nu}\gamma_5][\gamma_\nu) + i (\gamma_\nu\gamma_5][\sigma^{\mu\nu}) + i (\sigma^{\mu\nu}][\gamma_\nu\gamma_5 )  \biggr\},\\
      (\sigma^{\mu\nu})[\gamma_\nu] &=& \frac{1}{4} \biggl\{
      3i \biggl[ (\gamma^\mu][{\bf 1}) - ({\bf 1}][\gamma^\mu) - (\gamma^\mu\gamma_5][\gamma_5) - (\gamma_5][\gamma^\mu\gamma_5)\biggr]-(\sigma^{\mu\nu}][\gamma_\nu) \nonumber \\
      && - (\gamma_\nu][\sigma^{\mu\nu}) -(\sigma^{\mu\nu}\gamma_5][\gamma_\nu\gamma_5) + (\gamma_\nu\gamma_5][\sigma^{\mu\nu}\gamma_5)\biggr\},\\
      (\sigma^{\mu\nu})[\gamma_\nu\gamma_5] &=& \frac{1}{4}\biggl\{
      (\gamma_\nu][\sigma^{\mu\nu}\gamma_5) - (\sigma^{\mu\nu} \gamma_5][\gamma_\nu) - (\gamma_\nu \gamma_5][\sigma^{\mu\nu}) - (\sigma^{\mu\nu}][\gamma_\nu\gamma_5) \nonumber \\
        && + 3 i \biggl[ (\gamma^\mu \gamma_5][{\bf 1}) - ({\bf 1}][\gamma^\mu\gamma_5) - (\gamma^\mu][\gamma_5) - (\gamma_5][\gamma^\mu)\biggr]\biggr\},\\
      (\gamma_\mu)[\sigma^{\mu\nu}\gamma_5] &=& \frac{1}{4} \biggl\{
      3 i \biggl[ (\gamma^\nu][\gamma_5) - (\gamma_5][\gamma^\nu) + (\gamma^\nu\gamma_5][{\bf 1}) + ({\bf 1}][\gamma^\nu\gamma_5) \biggr] - (\gamma_\mu][\sigma^{\mu\nu}\gamma_5) \nonumber \\
      &&- (\sigma^{\mu\nu}\gamma_5][\gamma_\mu) -(\gamma_\mu\gamma_5][\sigma^{\mu\nu}) + (\sigma^{\mu\nu}][\gamma_\mu\gamma_5)\biggr\},\\      
      (\sigma^{\mu\alpha})[\sigma_\alpha^{\,\,\,\nu}] &=& (\sigma^{\nu\alpha})[\sigma_\alpha^{\,\,\,\mu} ]+ i (\sigma^{\mu\nu} ][{\bf 1}) - i ({\bf 1}][\sigma^{\mu\nu}) +i (\sigma^{\mu\nu}\gamma_5][\gamma_5) -i (\gamma_5][\sigma^{\mu\nu} \gamma_5),\\
      (\gamma_5)[\sigma^{\mu\nu}] &=& \frac{1}{4} \biggl\{
      i (\gamma^\mu \gamma_5][\gamma^\nu) -i (\gamma^\nu\gamma_5][\gamma^\mu) - i (\gamma^\mu][\gamma^\nu \gamma_5) + i (\gamma^\nu][\gamma^\mu \gamma_5) + (\sigma^{\mu\nu}][\gamma_5)  \nonumber \\
      && + (\gamma_5][\sigma^{\mu\nu}) + (\sigma^{\mu\nu} \gamma_5][{\bf 1}) + ({\bf 1}][\sigma^{\mu\nu}\gamma_5)+ i (\sigma^{\mu\alpha}\gamma_5][\sigma_\alpha^{\,\,\,\nu}) - i (\sigma^{\nu\alpha}\gamma_5][\sigma_{\alpha}^{\,\,\,\mu}) \nonumber \\
      &&
        -\epsilon^{\mu\nu\alpha\beta} (\gamma_\alpha][\gamma_\beta) + \epsilon^{\mu\nu\alpha\beta} (\gamma_\alpha\gamma_5][\gamma_\beta \gamma_5) \biggr\},\\
        (\gamma^\mu)[\gamma^\nu\gamma_5] &=& (\gamma^\nu)[\gamma^\mu\gamma_5] + \frac{i}{2} \biggl\{ (\sigma^{\mu\nu}][\gamma_5) + (\gamma_5][\sigma^{\mu\nu}) - (\sigma^{\mu\nu} \gamma_5][{\bf 1}) - ({\bf 1}][\sigma^{\mu\nu} \gamma_5) \nonumber \\
      &&+ \epsilon^{\mu\nu\alpha\beta} (\gamma_\alpha][\gamma_\beta) + \epsilon^{\mu\nu\alpha\beta}(\gamma_\alpha \gamma_5][\gamma_\beta \gamma_5)\biggr\}.
        \end{eqnarray}
}        
Together with further isospin Fierz relations,
\begin{eqnarray}
  ({\vec{\tau}})[{\bf 1}] + ({\bf 1})[{\vec{\tau}}]= ({\vec{\tau}}][{\bf 1}) + ({\bf 1}][{\vec{\tau}}),\\
      ({\vec{\tau}})[{\bf 1}] - ({\bf 1})[{\vec{\tau}}]=-i ({\vec{\tau}}]\times [{\vec{\tau}}),
\end{eqnarray}
one can establish the following linear relations,
{\allowdisplaybreaks
  \begin{eqnarray}
    O_3&=&-\frac{1}{4} \biggl(  O_3 + \frac{1}{4} O_{19} + \frac{1}{2} O_5 + \frac{1}{4} O_{20} \biggr),\\
    O_4&=&-\frac{1}{4} \biggl(  O_4 + \frac{1}{2} O_5 - i O_{15} + \frac{1}{4}O_{21} \biggr),\\
    O_5&=&-\frac{1}{4}\biggl(3  O_3 +  O_4 + i O_{15} + \frac{3}{4} O_{19} -\frac{1}{4} O_{20} + \frac{1}{4} O_{21} \biggr),\\
    O_6 &=& -\frac{1}{4} \biggl(O_6 + \frac{1}{2} O_8 + i O_{22} + i O_{23} \biggr),\\
    O_7 &=& -\frac{1}{4} \biggl(O_7 + O_8 -i O_{15} + O_{16} +i O_{24} \biggr),\\
    O_8 &=& -\frac{1}{4} \biggl( 3 O_6 + O_7 + i O_{15} - O_{16} + 3 i O_{22} - i O_{23} + i O_{24} \biggr),\\
    O_9 &=& -\frac{1}{4} \biggl(O_9 + \frac{1}{2} O_{11} - i O_{22} - \frac{i}{2} O_{24} + \frac{i}{2} O_{25} \biggr),\\
    O_{10} &=& -\frac{1}{4} \biggl( O_{10} + \frac{1}{2} O_{11} + i O_{15} + O_{17} - i O_{23} -\frac{i}{2} O_{24} - \frac{i}{2} O_{25} \biggr),\\
    O_{11} &=& -\frac{1}{4} \biggl(3 O_9 + O_{10} - i O_{15} - O_{17} - 3 i O_{22} - i O_{23} -i O_{25} \biggr), \\
    O_{12} &=& -\frac{1}{4} \biggl( O_{12} + \frac{1}{2} O_{14} + O_{19} + \frac{1}{2} O_{21} + 2 i O_{22} + i O_{24} + i O_{25} \biggr),\\
    O_{13} &=& -\frac{1}{4} \biggl( O_{13} + \frac{1}{2} O_{14} + O_{16} - O_{17} - O_{18} + O_{20} +\frac{1}{2} O_{21} + 2 i O_{23} + i O_{24} \nonumber \\
    && - i O_{25} \biggr),\\
    O_{14} &=& -\frac{1}{4} \biggl( 3 O_{12} + O_{13} - O_{16} + O_{17} + O_{18} + 3 O_{19} + O_{20} + 6 i O_{22} + 2 i O_{23} \nonumber \\
    && - 2 i O_{25} \biggr),\\
    O_{15} &=& -\frac{1}{4} \biggl( i O_{11} - 2 i O_{10} +2 O_{23} - O_{24} - O_{25} \biggr),\\
    O_{16} &=& -\frac{1}{4} \biggl( 6 O_7 - 3 O_8 - 2 i O_{25}\biggr),\\
    O_{17} &=& -\frac{1}{4} \biggl( 6 O_{10} - 3 O_{11} - 2 i O_{23} + i O_{24} + i O_{25} \biggr),\\
    O_{18} &=& O_{13} - \frac{1}{2} O_{14} + O_{20} - \frac{1}{2} O_{21} ,\\
    O_{19} &=& -\frac{1}{4} \biggl( O_{12} + \frac{1}{2} O_{14} + O_{19} + \frac{1}{2} O_{21} - 2 i O_{22} - i O_{24} - i O_{25} \biggr),\\
    O_{20} &=& -\frac{1}{4} \biggl(O_{13} + \frac{1}{2} O_{14} - O_{16} + O_{17} - O_{18} + O_{20} + \frac{1}{2} O_{21} - 2 i O_{23} - i O_{24}  \nonumber \\
    &&  + i O_{25} \biggr),\\
    O_{21} &=& -\frac{1}{4} \biggl( 3 O_{12} + O_{13} + O_{16} - O_{17} + O_{18} + 3 O_{19} + O_{20} - 6 i O_{22} - 2 i O_{23}  \nonumber \\
    &&  + 2 i O_{25} \biggr),\\
    O_{22} &=& \frac{i}{4} \biggl(O_{12} + \frac{1}{2} O_{14} - O_{19} - \frac{1}{2} O_{21} \biggr),\\
    O_{23} &=& \frac{i}{4} \biggl(O_{13} + \frac{1}{2} O_{14} - O_{16} - O_{17} - O_{20} - \frac{1}{2} O_{21} \biggr),\\
    O_{24} &=& \frac{i}{4} \biggl( 3 O_{12} + O_{13}+ O_{16} + O_{17} - 3 O_{19} - O_{20} \biggr),\\
    O_{25} &=& \frac{i}{4} \biggl(3 O_{12} - O_{13} - O_{14} - O_{16} - O_{17} - 3 O_{19} + O_{20} + O_{21} \biggr),\\
    O_{15} &=& -\frac{i}{4} \biggl( 2 O_7 - O_8 + 2 i O_{25} \biggr),\\
    O_{22} &=& -\frac{1}{4} \biggl(3 i O_9 + \frac{3}{2} i O_{11} - O_{22} -\frac{1}{2} O_{24} + \frac{1}{2} O_{25} \biggr),\\
    O_{23} &=& -\frac{1}{4} \biggl(3 i O_{10} +\frac{3}{2} i O_{11} + 3 O_{15} + i O_{17} - O_{23} - \frac{1}{2} O_{24} - \frac{1}{2} O_{25} \biggr),\\
    O_{24} &=& -\frac{1}{4} \biggl(9 i O_9 + 3 i O_{10} - 3 O_{15} - i O_{17} - 3 O_{22} - O_{23} - O_{25}\biggr),\\
    O_{25} &=& -\frac{1}{4} \biggl(-9 i O_9 + 3 i O_{10} +3 i O_{11} - 3 O_{15} - i O_{17} + 3 O_{22} - O_{23} - O_{24}\biggr).
  \end{eqnarray}
  }
By examining the above relations one finds that all operators can be expressed as linear combinations of the $O_X$, $X=S,P,V,A,T$, defined in Eq.~(\ref{eq:relp0}).

\end{document}